\documentclass[twocolumn]{aastex631}
\usepackage{CJK}
\usepackage{relsize}
\newcommand\BD{BD+60~1417B} 
\definecolor{red2}{rgb}{0.89, 0.02, 0.17}
\newcommand\WISE{W0047}

\usepackage[autostyle]{csquotes}
\usepackage{subfiles}
\usepackage{soul}
\usepackage{float}
\graphicspath{{./}{Plots}}
\usepackage{amsmath}
\usepackage{soul}
\usepackage{multirow}
\usepackage{booktabs}
\usepackage{hyperref}
\usepackage{graphics}

\received{April 26, 2024}
\revised{June 14, 2024}
\accepted{June 28 2024}

\shortauthors{Phillips et al.}
\graphicspath{{./}{Plots/}}

\newcommand{\teff}{T_{\rm eff}}
\newcommand{\logg}{{\rm log}(g)}
\newcommand{\mh}{[{\rm M/H}]}
\newcommand{\am}{[\alpha/{\rm M}]}
\newcommand{\vmicro}{v_{\rm micro}}
\newcommand{\vmacro}{v_{\rm macro}}
\newcommand{\vsini}{v\sin(i)}

\begin{document}
\title{Retrieving Young Cloudy L-Dwarfs: A Nearby Planetary-Mass Companion \BD~and Its  Isolated Red Twin W0047}
\shorttitle{Cloudy L-Dwarfs are Hard}
\begin{CJK*}{UTF8}{gbsn}
\shortauthors{Phillips et al.}
\correspondingauthor{Caprice Phillips}
\email{phillips.1622@buckeyemail.osu.edu}
\author[0000-0001-5610-5328]{Caprice L. Phillips}
\affil{Department of Astronomy, The Ohio State University, 100 W 18th Ave, Columbus, OH 43210 USA}

\author[0000-0001-6251-0573]{Jacqueline K. Faherty}
\affiliation{Department of Astrophysics, American Museum of Natural History, New York, NY 10024, USA}

\author[0000-0003-4600-5627]{Ben Burningham}
\affil{Centre for Astrophysics Research, School of Physics, Astronomy and Mathematics, University of Hertfordshire, Hatfield AL10 9AB}

\author[0000-0003-0489-1528]{Johanna M. Vos}
\affiliation{School of Physics, Trinity College Dublin, The University of Dublin, Dublin 2, Ireland}

\author[0000-0003-1445-9923]{Eileen Gonzales}
\altaffiliation{51 Pegasi b Fellow}
\affil{Department of Physics and Astronomy, San Francisco State University, 1600 Holloway Ave., San Francisco, CA 94132, USA}
\affil{Department of Astronomy and Carl Sagan Institute, Cornell University, 122 Sciences Drive, Ithaca, NY 14853, USA}

\author[0000-0001-9345-9977]{Emily J. Griffith}
\altaffiliation{NSF Astronomy and Astrophysics Postdoctoral Fellow}
\affil{Center for Astrophysics and Space Astronomy, Department of Astrophysical and Planetary Sciences, University of Colorado, 389 UCB, Boulder, CO 80309-0389}

\author[0000-0003-0548-0093]{Sherelyn Alejandro Merchan}
\affil{Department of Astrophysics, American Museum of Natural History, New York, NY 10024, USA}
\affil{ Hunter College, City University of New York, 695 Park Avenue, New York, NY 10065, USA}

\author[0000-0002-2682-0790]{Emily Calamari}
\affil{The Graduate Center, City University of New York, New York, NY 10016, USA}
\affil{Department of Astrophysics, American Museum of Natural History, New York, NY 10024, USA }

\author[0000-0001-6627-6067]{Channon Visscher}
\affiliation{Chemistry \& Planetary Sciences, Dordt University, Sioux Center IA 51250}

\affiliation{Center for Extrasolar Planetary Systems, Space Science Institute, Boulder, CO 80301}

\author[0000-0002-4404-0456]{Caroline V. Morley}
\affil{Department of Astronomy, The University of Texas at Austin, Austin, TX 78712, USA}

\author[0000-0001-8818-1544]{Niall Whiteford}
\affil{Department of Astrophysics, American Museum of Natural History, New York, NY 10024, USA }

\author{Josefine Gaarn}
\affil{3 Centre for Astrophysics Research, School of Physics, Astronomy and Mathematics, University of Hertfordshire, Hatfield AL10 9AB, UK}

\author[0000-0002-0551-046X]{Ilya Ilyin}
\affil{Leibniz-Institut for Astrophysics Potsdam (AIP), An der Sternwarte 16, D-14482 Potsdam, Germany}

\author[0000-0002-6192-6494]{Klaus Strassmeier}
\affil{Leibniz-Institut for Astrophysics Potsdam (AIP), An der Sternwarte 16, D-14482 Potsdam, Germany}

\begin{abstract}
We present an atmospheric retrieval analysis on a set of young, cloudy, red L-dwarfs -- CWISER J124332.12+600126.2 and WISEP J004701.06+680352.1 -- using the \textit{Brewster} retrieval framework. We also present the first elemental abundance measurements of the young K-dwarf (K0) host star, BD+60 1417 using high resolution~(R = 50,000) spectra taken with PEPSI/LBT. In the complex cloudy L-dwarf regime the emergence of condensate cloud species complicates retrieval analysis when only near-infrared data is available. We find that for both L dwarfs in this work, despite testing three different thermal profile parameterizations we are unable to constrain reliable abundance measurements and thus the C/O ratio. While we can not conclude what the abundances are, we can conclude that the data strongly favor a cloud model over a cloudless model.  We note that the difficulty in retrieval constraints persists regardless of the signal to noise of the data examined (S/N $\sim$ 10 for CWISER J124332.12+600126.2 and~40 for WISEP J004701.06+680352.1).  The results presented in this work provide valuable lessons about retrieving young, low-surface gravity, cloudy L-dwarfs. This work provides continued evidence of missing information in models and the crucial need for JWST to guide and inform retrieval analysis in this regime.
\end{abstract}


\section{Introduction}
\label{sec:introduction}

Brown dwarfs are a class of astronomical objects that are not massive enough~(13 -- 75 M$_{\mathrm{Jup}}$) to stably fuse hydrogen in their cores \citep{Kumar63}.  As such, they contract and cool as they age and progress through spectral types \citep{Kirkpatrick2005}. As the bridge between low-mass stars and directly-imaged gas-giants, brown dwarfs are key to understanding planetary atmospheres and planetary formation mechanisms~\citep{Faherty2016,Liu2016}. Brown dwarfs that share the same parameter space (e.g., wide orbits/long orbital periods, colors, mass, radius, temperature, and spectral morphology) with directly-imaged planets can be complementary to study~\citep{Burgasser2010, Faherty2013, Faherty2013BD,Faherty2021}. Furthermore, young, low mass brown dwarfs are valuable laboratories for refining models of atmospheric chemistry and cloud formation. Studies have shown evidence for a correlation between clouds and youth~\citep{Faherty2013,Faherty2016,Vos2018}. At younger ages (low surface gravities), the sedimentation and appearance of clouds are affected, further highlighting the value of studying young, low mass brown dwarfs~\citep{Marley2012, Suarez2022,Suarez2023}. As analogs of young directly-imaged exoplanets they are ideal objects to test and verify models and atmospheric retrieval techniques.  The direct detection and atmospheric studies of exoplanets remains a challenge due to the high contrast ratio with the host star~\citep{HR8799Nature,Chilcote2017,Rajan2017,Wang2020,RuffioHR8799, Whiteford2023}. Brown dwarfs can be studied in far greater detail than directly-imaged exoplanets, as they are easier to observe. Overall, brown dwarfs have been used to inform our observational knowledge of extrasolar atmospheres and advance theoretical understanding of cool worlds~\citep{Burrows2011,Tremblin2017,Marley2021}.

\par

As far back as the discovery of the first methane bearing brown dwarf, Gliese 229B (\citealt{oppenheimer1995}; \citealt{nakajima1995}) the exact formation mechanism for brown dwarfs has been debated. While they are largely thought to form like stars via cloud fragmentation, there are alternative pathways which might lead to the lowest mass isolated sources~\citep{Bate2002,Whitworth2005,Whitworth2006,Bonnell2008}.   Disk fragmentation and core accretion are two formation mechanisms that overlap with giant, widely separated exoplanets. Distinguishing between formation pathways has been tackled on the theoretical side (e.g., \citealt{Whiteford2007}; \citealt{Luhman2007}; \citealt{Whiteworth2010}; \citealt{Whitworth2018}) and recent space-based and ground-based data allows us to address it on the observational side with carbon-to-oxygen (C/O), [C/H] and [O/H] ratios (e.g., \citealt{Madhu2016}; \citealt{Maldonado2017}; \citealt{Gonzales2020}; \citealt{Burningham2021}; \citealt{Gonzales2022}; \citealt{Calamari2022}; \citealt{Vos2023}).
 \par
C/O ratios may help inform what pathway of formation a given object followed, shedding light on whether brown dwarfs form in a disk like planets in our solar system or in a fragment of a molecular cloud like binary stars~\citep{Oberg2011}. For example, at large orbital radii, we expect particularly strong differential evolution between gas and dust in the disk, which may translate to large observational differences in composition between substellar objects and their primary stars. Pertaining to planets, a stellar C/O ratio is expected for objects formed through gravitational instability, while super-stellar C/O indicates accretion of solids. The C/O ratio of the companion can be compared to the host star to provide context for formation and evolution history. With spectroscopic observations of a host star, we can measure the star's photospheric abundances -- a chemical fingerprint of the gas the system was born out of and then have the base level for comparison to a companion~\citep{Konopacky2013,Wang2022,Xuan2023, Hoch2023}. 
\par

Exoplanetary atmospheric retrieval (or the spectral retrieval method) refers to the inference of atmospheric properties of an exoplanet given an observed spectrum \citep{Madhu2019}. Using this method, one can retrieve the atmospheric abundances and in turn derive bulk abundances which can be compared to their host star values and extract clues to their formation mechanisms (e.g., \citealt{Wang2022}; \citealt{Wang2023}; \citealt{Xuan2022}).

While a suite of retrieval frameworks have been developed in recent years (see review from \citealt{MacDonald2023}), for this work, we use the \textit{Brewster}~\citep{Burningham2017} retrieval framework. \textit{Brewster} is optimized for cloudy L-dwarfs and was initially designed to model and capture the complexity in clouds in extrasolar atmospheres. \textit{Brewster} has been utilized and tested on brown dwarfs in the L - T spectral type regimes (e.g. \citealt{Burningham2017}; \citealt{Burningham2021}; \citealt{Gonzales2020}; \citealt{Gonzales2021}; \citealt{Gonzales2022}; \citealt{Calamari2022}; \citealt{Vos2023}; \citealt{Gaarn2023}).

In this paper, we employ the \textit{Brewster} spectral retrieval framework to understand the atmospheric structure and composition of two young red L-dwarfs. In Section \ref{sec:literature_data}, we present literature data on both systems. In Section \ref{sec:data}, we present spectroscopic data for CWISER
J124332.12+600126.2~(hereinafter \BD)~and WISEP J004701.06+680352.1~(hereinafter \WISE)~along with newly obtained data for the host star BD+60 1417. In Section \ref{sec:SED}, we discuss SED analysis used in the \textit{Brewster} framework. We discuss the host star parameters and abundance analysis in Section \ref{sec:host_star_analysis}. In Section \ref{sec:brewster}, we discuss the retrieval framework and modification settings used in our work. We present the retrieval analysis results for \BD~and \WISE~in Sections \ref{sec:BD60_retrieval_results} and \ref{sec:W0047_retrievals} respectively. We provide key discussions in Section \ref{sec:discussion}, and conclude in Section \ref{sec:summary_conclusion}. 
\section{Young Red  L-Dwarf Spectroscopic Twins: \BD~and \WISE}
\label{sec:literature_data}

 The  nearby ($\sim$45 pc) recently discovered L6 -- L8$\gamma$ companion, \BD, is a planetary-mass (15 $\pm$ 5 M$_{\mathrm{Jup}}$) object prime for atmospheric characterization \citep{Faherty2021}. \BD~lies at a wide separation from its young~(50 -- 150 Myr) K0 host star ($\sim$ 1662 AU). \cite{Faherty2021} found that this object lies in sparsely populated region of mass ratio vs separation parameter space, with only 8 other objects, with an estimated mass $<$ 20 M$_{\mathrm{Jup}}$ and an orbital separation $>$ 1000 AU. \BD~has a mass  ratio in line with known exoplanets but appears too widely separated to have formed as a planet \emph{in situ}. Thus there are open questions regarding its formation history and mechanism.  
\par

\WISE~\citep{Gizis2015} is an isolated red L7$\gamma$ dwarf member of the $\sim$120Myr AB Doradus moving group. \cite{Faherty2021} found that it was a spectroscopic twin to \BD. \BD~and \WISE~have similar masses, temperatures, and extremely red colors of (J-K$_{s}$) = 2.72 $\pm$ 0.34 and (J-K) = 2.55 $\pm$ 0.2  respectively (see Table \ref{tab:BD60vsW0047}).  Additionally, both objects show signatures of condensate cloud driven variability~(e.g. \citealt{Vos2018}, Vos et al. in prep). \textit{Spitzer} photometric monitoring was obtained for \WISE, which showed high amplitude variability of 1.07 $\pm$ 0.04$\%$. This variability measurement put \WISE~amongst the highest \textit{Spitzer} variability amplitudes detected~\citep{Vos2018}. These results are consistent with near-IR variability amplitudes found for \WISE~using HST~(11$\%$)~\citep{Lew2016}. The similarities between \WISE~and \BD~therefore drives our intercomparison of their resultant retrieved properties.

\section{Data}
\label{sec:data}
\subsection{Spectroscopic Data for BD+60 1417}
\begin{figure*}
    \centering
    \includegraphics[width=\textwidth]{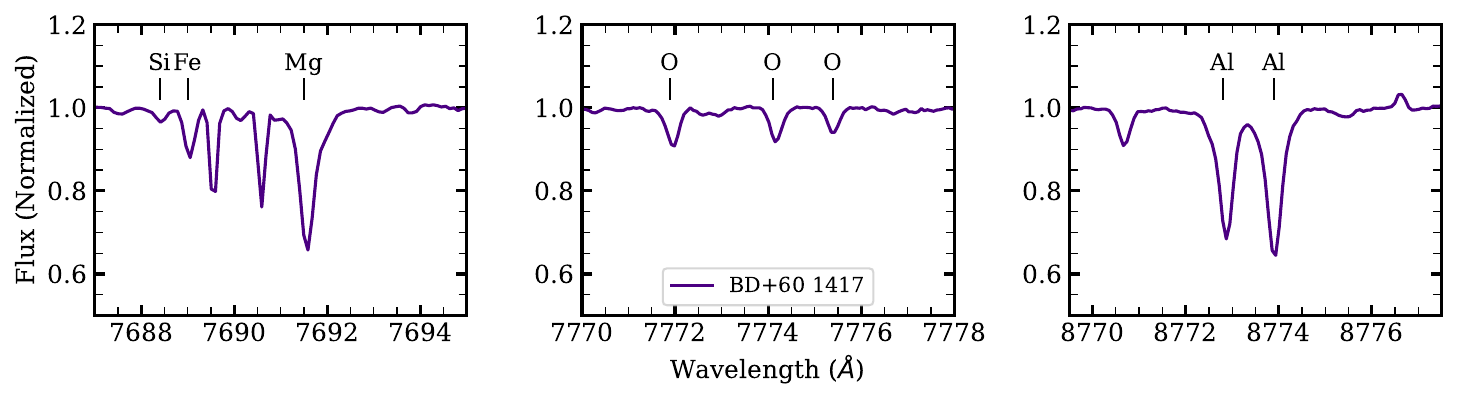}
    \caption{PEPSI spectra of BD+60 1417 in selected wavelength regions of $7687-7695\AA$ (left), $7770-7778\AA$ (center), and $8770-8777\AA$ (right). Line features used in the abundance analysis have been labeled.}
    \label{fig:host_spectra}
\end{figure*}

BD+60 1417 is a nearby ($\sim$ 45 pc; \citealt{GaiaDR3}) solar type star (K0 spectral type). As summarized in \citet{Faherty2021}, BD+60 1417 has numerous youth indicators including: a Li I abundance, measurement of a fast rotation rate, a position on color-magnitude and color-color diagrams indicating youth all of which point to an estimated age range of 50 -- 150 Myr.  However, until this paper, BD+60 1417 lacked a high resolution spectrum used to calculate solar abundances.  Therefore, we obtained high-resolution optical spectra of BD+60 1417 (V = 9.364 $\pm$ 0.11) from the Potsdam Echelle Polarimetric and Spectroscopic Instrument (PEPSI; \citealt{Strassmeier2015}) on the Large Binocular Telescope (LBT) on 9th February 2023. Observations were taken with the 300 $\mu$m fiber for a resolution $R = \lambda/\Delta\lambda =  50 000$ and with cross-dispersers (CD) III and VI to obtain high signal to noise spectra in the wavelength ranges 4800 -- 5411 and 7419 -- 9140~\AA.  This wavelength regime and resolution contains multiple stellar absorption features for C, O, Mg, Si, Ca, and Al. BD+60 1417 was observed for 1260 s, reaching a signal-to-noise ratio (S/N) of 795 in CD III and and 517 in CD VI. We plot selected regions of the spectra in Figure~\ref{fig:host_spectra}.

We employ the spectroscopic data systems for PEPSI pipeline (SDS4PEPSI) to reduce the spectra of BD+60 1417, as described in \citet{Strassmeier2018}. In short, this pipeline applies bias and flat-field corrections and estimates the photon-noise contribution to the spectra. It removes contamination from scattered light as well as cosmic rays. The pipeline conducts flux extraction and wavelength calibrates the spectra against Th-Ar lines, as well as corrects for the echelle's blaze function, fringing, and vignetting effects. Finally, the pipeline corrects for the star's radial velocity shifts to bring the spectra to rest-frame wavelengths and applies a continuum normalization by fitting a 2D smoothing spline in the cross-dispersion and dispersion directions.

\subsection{Spectroscopic Data for \BD~and \WISE}
For \BD, we use the published 1 -- 2.5 $\mu$m Infrared Telescope Facility (IRTF)/SpeX Prism spectrum from \cite{Faherty2021} with R $\sim$ 100 and  S/N $\sim$ 9. The spectra for \WISE~consist of a 1 -- 2.5 $\mu$m IRTF/SpeX spectrum from \cite{Gizis2012}  with R $\sim$ 150 and S/N $\sim$ 37 and a 1.10 -- 1.69 $\mu$m HST/WFC3 spectrum from \cite{Manja2019} with R $\sim$ 130 and S/N $\sim$ 370 at 1.25 $\mu$m.

\section{Fundamental Parameters from Spectral Energy Distributions for \BD~and \WISE~}
\label{sec:SED}
\begin{figure*}
    \centering
    \includegraphics[width=\textwidth]{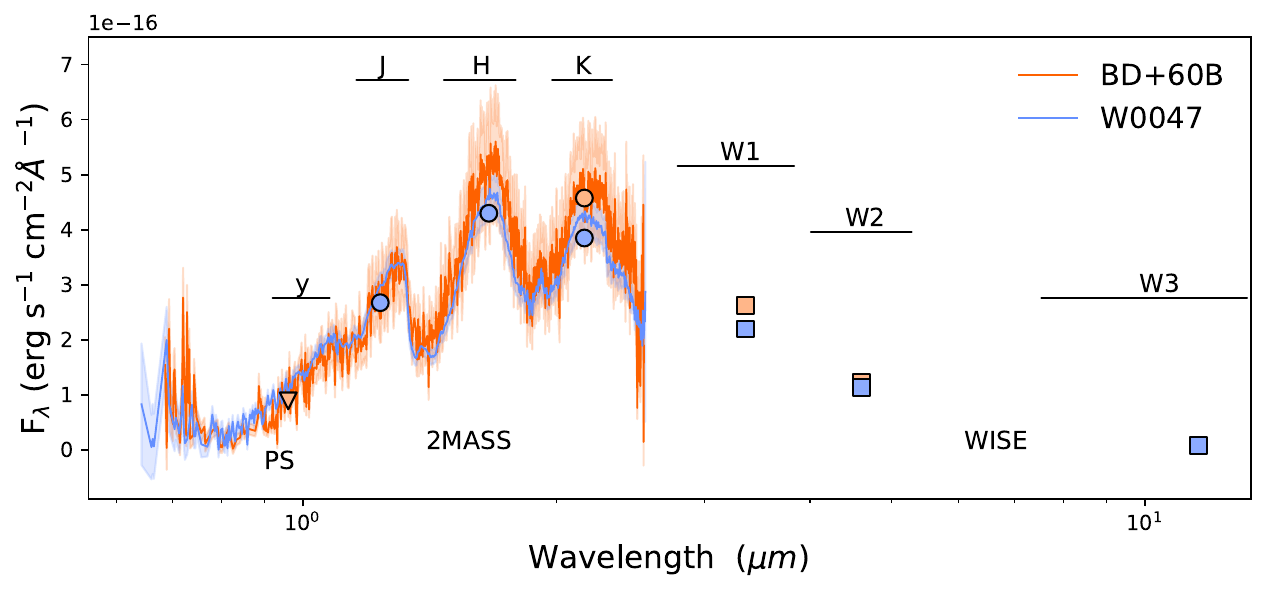}
    \caption{SEDs of \BD~and \WISE. The distance-calibrated SEDs of \BD~and \WISE~are shown in orange and blue, respectively. The SpeX/Prism spectrum for \BD~is shown in orange, with a shaded region indicating the flux uncertainties. Similarly for \WISE~the SpeX/Prism and HST/WFC3 spectrum are shown in blue. Photometric points for \WISE~are shown in a light blue  and in light orange for \BD. Circles are used for 2MASS, squares for WISE, and a down triangle for Pan-STARRS. The wavelength coverage for the photometric filters are indicated with horizontal lines.  Error bars on the photometric points are smaller than the point size.}  
    \label{fig:SED_BD_W0047}
\end{figure*}

In order to acquire fundamental parameters, we constructed and analyzed the SEDs for \BD~ and \WISE. We compiled all relevant data including the parallax measurements of \BD~and \WISE~along with spectra and photometry to acquire a distance-calibrated SED. WISE W1 and W2 photometry was used from the catWISE2020 reject table, as \BD~was rejected from the main table because it was near a diffraction spike from the primary. We follow \cite{Faherty2021} and find the photometry to  be usable for construction and analysis of the SED.  We used the open-source package SEDkit \citep{Filippazzo20} presented in \citet{Filippazzo15} to construct the spectral energy distribution (SED) and derive the fundamental parameters of these twin analogs. Integrating under the flux calibrated SED, we calculated the $L_{\mathrm{bol}}$ for each target. Pairing the $L_{\mathrm{bol}}$ values and ages of $100\pm50$ Myr \citep{Faherty2021}  and $120\pm50$ Myr \citep{Gizis2015} for \BD~and \WISE, respectively. Along with the \citet{Marley2021} evolutionary models, we semi-empirically calculate the $T_{\rm eff}$ and infer the radius, mass and $\log g$. The spectral energy distribution of \BD~and \WISE~are shown in Figure \ref{fig:SED_BD_W0047} and fundamental parameters derived from the SEDs are shown in Table \ref{tab:BD60vsW0047}. 

The SED study and derived fundamental parameters for \BD~and \WISE~emphasize the similar nature of these sub-stellar objects, as indicated by their $T_{\rm eff}$ being within $\sim 50$K, identical radii and mass measurements that extend into the nominal space between brown dwarfs and planets.

\begin{deluxetable}{lrrc}[h!]
\label{tab:BD60vsW0047}
\tabletypesize{\small}
\tablecolumns{10}
\tablecaption{Observed properties and derived fundamental parameters for \BD~and \WISE}
\tablewidth{0pt}
\setlength{\tabcolsep}{0.025in}
\tablehead{\colhead{Property}& \colhead{\BD}  & \colhead{\WISE }&\colhead{Refs}}
\startdata
RA$^{a}$& 12:43:32.1025&00:47:00.3866 &C03\\
Dec$^{a}$ &+60:01:26.5150& +68:03:54.3672&C03   \\
$\mu_{\alpha}$ cos $\delta$ (mas yr$^{-1}$)&  $-133\pm8$    & $245.7 \pm 5.4$& M21, L16  \\
$\mu_{\delta}$ (mas yr$^{-1}$) &  $-55\pm8$&  -- 108.5 $\pm$ 2.4& M21, L16 \\ 
Spectral Type (IR) &L8$\gamma$&L7& F21, G15        \\
Distance (pc)& 44 $\pm$ 4 & 12.2 $\pm$ 0.4 & F21,G15\\
\hline
Photometry \\ 
\hline
Pan-STARRS y (mag)&20.481 $\pm$ 0.178&---&G21 \\
2MASS $J$ (mag) &          $>$17.452            & $15.604\pm0.068$   &   C03    \\
2MASS $H$ (mag) &    $>$16.745                   & $13.968\pm0.041$   &   C03     \\
2MASS $K_S$ (mag) &  $13.582\pm0.045$                   & $12.562\pm0.024$  &    C03    \\
WISE $W1$ (mag) &    $14.594\pm0.031$                   &  $11.876\pm0.023$  &  C12     \\
WISE $W2$ (mag) &    $13.872\pm0.035$                   &  $11.268\pm0.020$ &   C12     \\
WISE $W3$ (mag) & ---                      &  $10.327\pm0.072$  &  C13     \\
\hline
\multicolumn{3}{l}{Fundamental Parameters from SED Analysis}\\  
\hline
$L_{\mathrm{bol}}$ (L$_{bol}$/L{$_{\odot}$}) & $-4.42\pm0.10$&   $-4.35 \pm0.06$   &TW\\
$T_{\mathrm{eff}}$ (K)&   $1240\pm81$  &   $1291\pm51$  &   TW   \\
Radius $(R_{\mathrm{Jup}})$ & $1.29\pm0.06$   &  $1.29\pm0.03$  &   TW  \\
Mass  $(M_{\mathrm{Jup}})$  &  $13.47\pm5.67$             &   $15.65\pm4.65$   &     TW  \\
$\log g$ (cm s$^{-2}$) & $4.26\pm0.20$ & $4.3\pm0.1$ & TW        \\ 
\hline
\enddata
\tablenotetext{a}{Epoch J2000, ICRS}
\tablerefs{F21: \citet{Faherty2021}; TW: This Work; C03: \citet{Cutri2003}; C12: \citet{Cutri2012_data} G15: \citet{Gizis2015}, M21: \citet{Marocco2021}; L16: \citet{Liu2016}; G21: \citet{Gaia2021}}
\label{tab:props}
\end{deluxetable}

\section{Spectral Parameter and Abundance Determination for BD+60 1417}
\label{sec:host_star_analysis}

\subsection{Stellar Parameter Determination}
We used the PEPSI reduced spectrum to determine stellar parameters and stellar abundances with MOOG, a LTE radiative transfer code \citep{Sneden1973}, implemented through the iSpec wrapper \citep{BlancoCuaresma2014, BlancoCuaresma2019}.
Stellar parameters $\teff$, $\logg$, $\mh$, $\am$, and $\vmicro$ are fit with equivalent width analysis.
Through iSpec, we fit the Fe\,{\sc i} and Fe\,{\sc ii} lines in the Gaia-ESO version 6 line list \citep{Gilmore2012, Heiter2021} with Gaussian or Voigt profiles and measure their equivalent widths. We remove the strongest and weakest lines by requiring that the reduced equivalent width of the lines lie between -4.3 and -6.0 and by requiring that the lower-energy state of the lines lie between 0.5 and 5.0, as recommended by \citet{Mucciarelli2013}, \citet{BlancoCuaresma2014}, and \citet{BlancoCuaresma2019}. Using MARCS model atmospheres \citep{Gustafsson2008} and \citet{Asplund2009} solar abundances, we implement MOOG equivalent width parameter determination within iSpec, which iteratively fits models of $\teff$, $\logg$, $\mh$, $\am$, and microturbulence velocity ($\vmicro$), searching for a model that brings the Fe lines into ionization and excitation equilibrium. We initiate the parameter search at $\teff = 4993$ K, $\logg=4.55$ dex, and $\mh = 0.0$ dex, based on the stellar parameters in the TESS input catalog \citep{Stassun2018} adopted in \citet{Faherty2021}. iSpec returns best-fit stellar parameters of $\teff = 5382 \pm 164$ K, $\logg = 4.88\pm 0.06$ dex, $\mh = 0.37$ dex, $\am = 0.0$ dex, and $\vmicro = 1.026$ km/s.


In addition to the above stellar parameters, MOOG and iSpec require values of macroturbulence ($\vmacro$), rotational velocity ($\vsini$), spectral resolution ($R$), and limb-darkening coefficient to conduct stellar abundance determination with spectral synthesis. We set the resolution to that of the instrument ($R=50000$) and the limb-darkening coefficient to 0.6, as recommended by \citet{BlancoCuaresma2014}. Because the broadening terms $\vmacro$ and $\vsini$ are degenerate at $R=50000$, we set $\vmacro=0$ km/s and find the best fit $\vsini$ through a spectral synthesis fit to Fe\,{\sc i} and Fe\,{\sc ii} line windows with all other stellar parameters fixed to the values listed above \citet{Buder2021}. We find that a $\vsini=5.40$ km/s best minimizes the residual between the synthetic and observed spectra.

To calculate the error on our stellar parameters, we run the equivalent width and spectral synthesis fits described above at the upper and lower bounds of $\teff$ to find the correlated deviations in $\logg$, $\mh$, $\vmicro$, and $\vsini$. We adopt an error of 0.12 dex on $\logg$, 0.04 dex on $\mh$, $0.26$ km/s on $\vmicro$ and $0.35$ km/s on the broadening term $\vsini$. We list all best-fit stellar parameters and their errors in Table~\ref{tab:stellar_params}. 

We note that the photometric stellar parameters adopted as the initialization are not consistent with our calculated spectroscopic parameters. While the photometric $\teff$ and $\logg$ are derived from observables and used to calculate physical parameters such as stellar radii and mass, their adoption inhibits achieving ionization and excitation equilibrium in the Fe\,{\sc i} and Fe\,{\sc ii} abundances. We adopt the spectroscopic parameters in further analysis as they are optimized for stellar abundance determination, but stress that these values should not be used to derive the stellar mass or radius \citep[e.g.,][]{Jofre2019}.

\begin{deluxetable}{lr}[h!]
\tabletypesize{\small}
\tablecolumns{2}
\tablecaption{Stellar parameters and abundances for BD+60 1417}
\tablewidth{0pt}
\setlength{\tabcolsep}{0.3in}
\tablehead{\colhead{Property}               & \colhead{BD+60 1417   }}
\startdata
$\teff$     &  $5381 \pm 164$ K \\
$\logg$     & $4.88 \pm 0.12$ dex \\
$\mh$       & $0.37 \pm 0.04$ dex \\
$\am$       & 0.0 dex \\
$\vsini$    & $5.40 \pm 0.36$ km/s \\
$\vmicro$   & $1.03 \pm 0.26$ km/s \\\hline
$[{\rm C/H}]$       &   $-0.39 \pm 0.14$ dex  \\ 
$[{\rm O/H}]$       &   $0.00 \pm 0.18$ dex   \\
$[{\rm O/H}]_{\rm NLTE}$ & $-0.02 \pm 0.18$ dex \\
$[{\rm Mg/H}]$      &   $0.13 \pm 0.01$ dex   \\ 
$[{\rm Si/H}]$      &   $0.07 \pm 0.06$ dex   \\ 
$[{\rm Ca/H}]$      &   $0.36 \pm 0.02$ dex   \\ 
$[{\rm Al/H}]$      &   $0.15 \pm 0.03$ dex   \\ 
$[{\rm Fe/H}]$      &   $0.27 \pm 0.03$ dex   \\ \hline
C/O       &   $0.23 \pm 0.12$  \\ 
Mg/Si     &   $1.41 \pm 0.19$   \\ 
Ca/Al     &   $1.26 \pm 0.11$   \\ \hline
\enddata
\tablecomments{Unless noted, all [X/H] abundances are 1D LTE, relative to solar abundances in \citet{Asplund2009}. We provide LTE and NLTE [O/H] abundances. The reported C/O ratio is calculated from the NLTE [O/H] value.}
\label{tab:stellar_params}
\end{deluxetable}

\subsection{Stellar Abundance Determination}

We determine the stellar abundances of C, O, Mg, Si, Ca, Al, and Fe with MOOG spectral synthesis implemented through iSpec. We use a curated line list created from referencing line lists from \citet{fulbright2000}, GALAH DR3 \citep{Buder2021}, and Gaia-ESO \citep{Gilmore2012, Heiter2021}, and the solar spectrum \citep{moore1966} comprised of 5 \ion{C}{1} lines, 3 \ion{O}{1} lines, 16 \ion{Mg}{1} lines, 27 \ion{Si}{1} lines, 6 \ion{Ca}{1} lines, 4 \ion{Ca}{2} lines, 6 \ion{Al}{1} lines, 265 \ion{Fe}{1}, and 10 \ion{Fe}{2} lines. While the full spectrum was normalized in the Spectroscopic Data systems for PEPSI pipeline, we fit and normalize the continuum in the $8\AA$ region around each line window with a spline to ensure local continuum normalization.

 For each element, we create windows of $5\AA$ around all relevant line features fit for the elemental abundance, with iSpec iteratively creating synthetic stellar spectra and finding the template with the abundance value that minimizes the residual between the synthetic and observed spectrum. We further fit each line of each element individually, finding consistent abundances between the lines of different ionization states for Si and Fe. We report the best fit 1D LTE [C/H], [O/H], [Mg/H], [Si/H], [Ca/H], [Al/H], and [Fe/H] values in Table~\ref{tab:stellar_params}. We repeat the stellar abundance fits for each element with the upper and lower limits of the stellar parameters to calculate the abundance errors. The errors range from 0.01 dex (for Mg) to 0.18 dex (for O) and are also reported in Table~\ref{tab:stellar_params}.

 In this paper, we calculate 1D LTE abundances for BD+60 1417. 3D and non-LTE (NLTE) effects are known to be a source of abundance uncertainty and can often cause abundance corrections $>0.1$ dex \citep[e.g.][]{Jofre2019}. While we cannot integrate NLTE grids into iSpec, we use NLTE correction lookup tables from the Max Planck Institute for Astronomy (MPIA)\footnote{\url{https://nlte.mpia.de/gui-siuAC_secE.php}} to estimate the 1D NLTE corrections for \ion{O}{1} \citep{Bergemann2021}, \ion{Mg}{1} \citep{Bergemann2017}, \ion{Si}{1} \citep{Bergemann2013}, \ion{Ca}{1}, and \ion{Ca}{2} \citep{Mashonkina2007}. Tabulated NLTE corrections for C and Al are not yet available and are not accounted for in our reported errors. We find NLTE corrections of $<0.01$ dex for all available Mg and Si lines. While some Ca lines have significant NLTE corrections (-0.1 to 0.04 dex), the median value is $<0.01$ dex and does not impact our reported [Ca/H] abundance. The O triplet lines used in this analysis have a mean NLTE correction of -0.02 dex, leading to an NLTE [O/H] abundance of -0.02 dex.

From the [X/H] abundances, we can calculate the C/O, Mg/Si, and Ca/Al ratios. Each ratio represents the ratio of the number of atoms of element X to the number of atoms of element Y where $\text{X}/\text{Y} = N_{\rm X}/N_{\rm Y} = 10^{\log N_{\rm X}}/10^{\log N_{\rm Y}}$, and $\log N_{\rm X} = \log_{10}(N_{\rm X}/N_{\rm H}) + 12$. We calculate $\log N_{\rm X}$ as $[\text{X}/\text{H}] + \log N_{\rm X, \odot}$, using solar abundances $\log N_{\rm X, \odot}$ of 8.43 for C, 8.69 for O, 7.60 for Mg, 7.51 for Si, 6.34 for Ca, and 6.45 for Al \citep{Asplund2009}. We report the resulting ratios in Table~\ref{tab:stellar_params}.

\section{Brewster Framework For Retrieving \BD~and \WISE~}
\label{sec:brewster}
To extract comparative abundances for the brown dwarfs \BD~and \WISE, we used the atmospheric retrieval code called \textit{Brewster} \citep{Burningham2017}.  Below we give a short description of the \textit{Brewster} framework we utilized in this work however we encourage the reader to see a detailed description of \textit{Brewster} in \cite{Burningham2017, Burningham2021}.  Most notably, we differ from \cite{Burningham2017, Burningham2021} with the use of a version update~\citep{Calamari2022} that utilizes nested sampling via \texttt{PyMultiNest} \citep{Buchner2014} instead of \texttt{emcee}. 

\subsection{The Forward Model}
The \textit{Brewster} forward model uses the two-stream radiative transfer technique of \cite{Toon1989}, including scattering, as first introduced by \cite{McKay1989} and subsequently used by \citet{Marley1996, Saumon2008, Morley12}. We set up a 64 pressure layer (65 levels) atmosphere with geometric mean pressures from $\log P = -4$ to $2.4$ bars spaced in 0.1~dex intervals. 

\subsection{Thermal Profiles}

There are a range of thermal profiles used in retrieval analyses ranging from a five-parameters joint exponential power law~\citep{MadhuandSeager2009,Burningham2017} to a 17-parameters free profile~\citep{Line2015}. For this work, we aim to test a variety of thermal profiles for \BD~and \WISE~to determine the reliability of retrieved abundances, pressure-temperature (P-T) profiles, bulk properties from ground-based near-infrared data, and the effect different thermal profile parameterizations have on retrieved abundances due to the notorious difficulty to model low gravity objects~\citep[see thesis work by Niall Whiteford and\footnote{see \url{https://era.ed.ac.uk/handle/1842/39547?show=full}}][]{Nowak2020}.  A similar exploration was conducted by \cite{Gonzales2022} \& \cite{Rowland2023} for L-dwarfs in a different regime. 
\par

In this work we explore three different thermal profiles for our sources.
\begin{itemize}
    \item[$\blacksquare$]

\textbf{Case 1 -- Five point spline:} The initial thermal profile tested is a computationally simple five point parameterization in which we specify five temperature pressure points: the top ($T_{\mathrm{top}}$), bottom ($T_{\mathrm{bottom}}$) and middle of the atmosphere ($T_{\mathrm{middle}}$) and two midpoints between the top and middle ($T_{\mathrm{q1}}$), and bottom and middle ($T_{q3}$). These points
are calculated in order, beginning with $T_{\mathrm{bottom}}$, which is selected in range between zero and the maximum temperature
defined in our prior which is 4000 K in this work. Then $T_{\mathrm{top}}$ is chosen between zero and $T_\mathrm{{bottom}}$, $T_{\mathrm{middle}}$ chosen between $T_{\mathrm{top}}$ and $T_{\mathrm{bottom}}$, and the remaining two midpoints chosen between $T_{\mathrm{top}}$ and $T_{\mathrm{middle}}$ $T_{\mathrm{q1}}$, $T_{\mathrm{middle}}$ and $T_{\mathrm{bottom}}$ (T$_{q3}$). A uniform prior is assumed for each temperature
within its respective range. This does not allow for temperature inversions but can result in ``irregular” profiles.

    \item[$\blacksquare$] \textbf{Case 2 -- Lavie profile:} We introduce a second thermal profile parametrization developed and implemented by \cite{Lavie2017} in the HELIOS retrieval code and subsequently implemented into other retrieval frameworks~\citep{Whiteford2023}. The Lavie profile is a less flexible profile with a simpler parameterization consisting of two free parameters, $\kappa_{0}$ and $T_{\mathrm{int}}$. The profile used in \cite{Lavie2017}, originates from \cite{Heng2014}, using a reduced form of Equation 126: 
\begin{equation}
    T^{4} = \cfrac{T_{\mathrm{int}}^{4}}{4} \cdot \Big ( \cfrac{8}{3}+3 \Tilde{m}\kappa_{0}\Big)
\end{equation}

where $T_{\mathrm{int}}$ is the internal temperature and $\kappa_{0}$ the constant component of the infrared opacity. $\Tilde{m}$ is column density determined via $P_{0}$ = $\Tilde{m}$ $\cdot ~g$, where $g$ is the surface gravity. 

\par

 \item[$\blacksquare$] \textbf{Case 3 -- Molli\`ere profile:}  Lastly, we explore a modified version of the thermal profile parameterization implemented in (\citealt{Molliere2020}; hereafter Molli\`ere ``Hybrid" Profile).  In the Molli\`ere ``Hybrid" Profile, the atmosphere is split into three distinct regions: the photosphere (middle altitudes), high altitude, and troposphere (low altitudes). In the photosphere region, the temperature is set via an Eddington approximation: 
\begin{equation}
    T(\tau)^{4} = \cfrac{3}{4}T_{0}^{4} (\cfrac{2}{3}+\tau)
\end{equation}
where  $T_{0}$ is a free parameter and  $\tau = \delta P^{\alpha}$. For $\tau$, $\delta$ and $\alpha$ are free parameters as well. 

The high altitude region is similar to that of \cite{Molliere2020}. In the troposphere, our Molli\`ere ``Hybrid" Profile differs from the implementation of \cite{Molliere2020}, in that a dry adiabatic atmosphere (for H$_{2}$/He) is used instead of a moist adiabatic. For a more detailed description of the  Molli\`ere ``Hybrid" Profile we refer the reader to \cite{Molliere2020}.
\end{itemize}
\subsection{Gas Opacities}

We consider the following absorbing gases in our analysis:
H$_{2}$O, CO, CO$_{2}$, CH$_{4}$, CrH, FeH, Na, and K. From \cite{Kirkpatrick2005}, these gases are chosen as they have been previously identified as important absorbing species in the spectra of L6 - L8 spectral types. 
\par
Optical depths due to absorbing gases for each layer are calculated using opacities sampled at a resolving power R~$=10,000$ from the \cite{Freedman_2008,Freedman_2014} collection with the updated opacites from \cite{Burningham2017}. Line opacites in our pressure-temperature range are tabulated in 0.5~dex steps for pressure and for temperature in steps ranging from 20~K to 500~K as we move from 75~K to 4000~K, which is then linearly interpolated to our working pressure grid.

We include continuum opacities for H$_{2}$-H$_{2}$ and H$_{2}$-He collisionally induced absorption, using cross-sections from \citet{Richard12} and \citet{Saumon2012}, and include Rayleigh scattering due to H$_{2}$, He and CH$_{4}$ but we neglect the remaining gases. Free-free continuum opacities are included for H$^{-}$ and H$_2^{-}$ as well as bound-free continuum opacity for H$^{-}$ \citep{Bell87,Bell80,John88}.

\subsection{Determining Gas Abundances \label{sec:gas_abundances}} 
We use the vertically constant mixing ratios method to determine gas abundances. The vertically constant method is sometimes referred to as a  ``free'' retrieval method which enables us to directly retrieve individual gas abundances under a computationally simple assumption of a single abundance for each gas throughout the 1D modelled atmosphere. It is considered ``free'' in the sense that there are no constraints imposed by thermochemical self-consistency. However, the simplicity of this method can not capture important variations with altitude that are expected for some species which can vary by several orders of magnitude in the photosphere~\citep{Rowland2023}.

\subsection{Cloud Model \label{sec:cloudmodel}}
We use the same simple cloud model utilized in previous works (e.g., \citealt{Burningham2017}, \citealt{Gonzales2020}, and \citealt{Vos2023}) where the cloud is parameterized as a ``deck'' or ``slab''. Both cloud types are defined where opacity due to the cloud is distributed among layers in pressure space and the optical depth defined as either grey or as a power-law ($\tau = \tau_0\lambda^\alpha$, where $\tau_0$ is the optical depth at 1~$\mu$m). As done in \cite{Gonzales2020}, the single scattering albedo is set to zero thereby assuming an absorbing cloud.

The deck cloud is defined to always become optically thick at some pressure $\mathrm{top}$, such that we only see the cloud top and the vertical extent of the cloud at pressures lower than $P_{\mathrm{top}}$. There are three parameters for the deck cloud: (1) a cloud top pressure $P_{\mathrm{top}}$, the point at which the cloud passes $\tau=1$ (looking down) for the first time ($P_{deck}$), (2) the decay height $\Delta \log P$, over which the optical depth falls to lower pressures as $d\tau/dP \propto \exp((P-P_{deck}) / \Phi)$ where $\Phi = (P_{\mathrm{top}}(10^{\Delta \log P} - 1))/(10^{\Delta \log P})$, and (3) the cloud particle single-scattering albedo ($\alpha)$. At pressures $P >P_{\mathrm{top}}$, the optical depth increases following the decay function until $\Delta \tau_{layer} = 100$. Deep below the cloud top there is essentially no atmospheric information as the deck cloud can rapidly become opaque with increasing pressure. Therefore, we stress that the pressure-temperature (P-T) profile below the deck cloud top extends the gradient at the cloud top pressure. 

The slab cloud differs from the deck as it is possible to see the bottom of the slab. The slab parameters include deck parameters (1) and (3). Because it is possible to see to the bottom of the slab, parameter (2) is now a physical extent in log-pressure ($\Delta \log P$). We also include an additional parameter for determining the total optical depth of the slab at 1$\mu$m ($\tau_{\mathrm{cloud}}$), bringing the total number of parameters to 4. The optical depth is distributed through the slab's extent as $d\tau / dP \propto P$ (looking down), reaching its total value at the highest pressure (bottom) of the slab. The slab can have any optical depth in principle, however, we restrict our prior as $0.0 \leq \tau_{\mathrm{cloud}} \leq 100.0$.

If the deck or slab cloud is non-grey, an additional parameter for the power ($\alpha$) in the optical depth is included.

\section{Retrieval Model}

The retrieval process consists of optimizing the parameters
of the forward model such that the resultant spectrum provides
the best match to the observed spectrum. As described by
\cite{Burningham2017}, we use a Bayesian framework to optimize the model fit to the data by varying the input parameters. Following \cite{Calamari2022}, we use a modified sampler to the original version of \textit{Brewster}, that uses \texttt{PyMultinest}~\citep{Buchner2014} instead of \texttt{emcee}~\citep{Foreman-Mackey_2013}.

\par
\textit{Brewster} applies Bayes' theorem to calculate the ``posterier probability", $p(\textbf{x}|\textbf{y})$, the probability of a set of parameters' $(\textbf{x})$ truth value given some data $(\textbf{y})$, in the following way:
\begin{equation}
    p(\textbf{x}|\textbf{y}) = \frac{\mathcal{L}(\textbf{x}|\textbf{y})p(\textbf{x})}{p(\textbf{y})}
\end{equation}

where $\mathcal{L}(\textbf{x}|\textbf{y})$ is the likelihood that quantifies how well the data match the model, $p(\textbf{x})$ is the prior probability on the parameter set and $p(\textbf{y})$ is the probability of the data marginalized over all parameter values, also known as the Bayesian evidence.

 The posterior probability space is explored using the \texttt{PyMultiNest} sampler \citep{Buchner2014} which utilizes nested sampling to discover the set of parameters with the maximum likelihood given the data. The samples in an n-dimensional hypercube, or state-vector, are translated into parameter values via the ``prior-map". The prior-map function is how prior probabilities are set for each parameter and are then transformed into appropriate parameter values to be used in the forward model. This algorithm is equipped to handle a parameter space that may contain multiple posterior modes and/or degeneracies in moderately high dimensions.

\begin{deluxetable*}{lcl}
\tablecaption{Priors for \BD~and \WISE~retrieval models\label{tab:Priors}} 
\tablehead{\colhead{Parameter} &\colhead{} &\colhead{Prior}}
  \startdata
  gas volume mixing ratio && uniform, log $f_{gas} \geq -12.0$, $\sum_{gas} f_{gas} \leq 1.0$ \\
  thermal profile~($T_{\mathrm{bottom}}$, $T_{\mathrm{Top}}$, $T_{\mathrm{middle}}$, $T_{\mathrm{q1}}$, $T_{\mathrm{q3}}$) && uniform, $0~\mathrm{K} < T < 4000.0\, \mathrm{K}$\\ 
  scale factor ($R^2/D^2$) && uniform, $0.5\,R_\mathrm{Jup} \leq\, R\, \leq 2.5\,R_\mathrm{Jup}$ \\
  gravity (log\,$g$)&& uniform, $1\,M_\mathrm{Jup} \leq\; gR^2/G\; \leq 80\,M_\mathrm{Jup}$ \\
  cloud top\tablenotemark{a} && uniform, $-4 \leq \mathrm{log}\, P_{CT} \leq 2.3$\\ 
  cloud decay scale\tablenotemark{b} && uniform, $0< \mathrm{log}\,\Delta\, P_{decay}<7$\\
  cloud thickness\tablenotemark{c} && uniform, log\,$P_{CT} \leq\,$log $(P_{CT}+\Delta P)\, \leq2.3$\\
  cloud total optical depth at $1\mu$m && uniform,  $0.0 \leq \tau_{cloud} \leq 80.0$ \\
  single scattering albedo ($\omega_0$) && uniform, $0.0 \leq \omega_0 \leq 1.0$\\
  wavelength shift && uniform, $-0.01 < \Delta \lambda <0.01 \mu$m \\
  tolerance factor && uniform, log($0.01 \times min(\sigma_{i}^2)) \leq b \leq$ log$(100 \times max(\sigma_{i}^2)) $\\
 \enddata
  \tablenotetext{a}{For the deck cloud this is the pressure where $\tau_{cloud} = 1$, for a slab cloud this is the top of the slab.}
  \tablenotetext{b}{Decay height above the $\tau_{cloud} = 1.0$ level only for deck cloud.}
  \tablenotetext{c}{Thickness and $\tau_{cloud}$ only retrieved for slab cloud.}
\end{deluxetable*}
\subsection{Model Selection}

A variety of models were tested in our retrievals of \BD~and~\WISE. These models differed in their cloud prescriptions and temperature-pressure profile parameterizations, which are explained in Section \ref{sec:brewster}. In order to compare all of our retrievals, we  calculate  the logarithm of the Bayesian evidence (logEv), where the highest logEv is preferred. For each of the models tested, the  $\Delta$logEv is calculated by subtracting the logEv of the winning model of the Bayesian evidence.

We use the following significance intervals from \cite{KassandRaftery} to distinguish between two models, with evidence against the lower logEv as:
\begin{itemize}
    \item 0 $<$ $\Delta$logEv $<$ 0.5: no preference worth mentioning
    \item 0.5 $<$ $\Delta$logEv $<$ 1: positive
    \item 1 $<$ $\Delta$logEv $<$ 2: strong
    \item $\Delta$logEv $>$ 2: very strong
\end{itemize}
We began by building from the least complex model
(cloud-free) to the most complex (power law slab cloud model) for the three thermal profile parameterizations tested in this work for \BD~and~\WISE~(see Table \ref{tab:models_tested}).
\begin{table*}[t]
    \centering
    \begin{tabular}{l|ccccc}
    \hline \hline
    Models&N params&\BD&\WISE&\BD&\WISE\\
    \hline
    \textbf{5 point P-T}&&$\Delta$logEv&$\Delta$logEv&logEv&logEv\\
    
    \textbf{ parameterization} \\
    \hline
    Cloudless&16&--21.3&--4.52E+03&7.767E+03&5.159E+03\\
    Grey Deck Cloud& 18&--0.7&--19.75&7.787E+03&9.660E+03 \\
    Grey Slab Cloud& 19&--9.5&--22.82&7.778E+03&9.657E+03\\
    Power Law Deck Cloud&19&0.0&0.00&7.788E+03&9.680E+03 \\
    Power Law Slab Cloud& 20&--5.9&--11.62&7.782E+03&9.668E+03\\
    \hline
    \textbf{Lavie Profile}&\\
    \hline
     Cloudless&13&--69.3&--176.66&7.702E+03&9.463E+03\\
    Grey Deck Cloud& 15&0.0&--74.75&7.772E+03&9.565E+03 \\
    Grey Slab Cloud& 16&--5.59&--80.7&7.767E+03&9.559E+03\\
    Power Law Deck Cloud&16&--6.5&0.00&7.766E+03&9.640E+03\\
    Power Law Slab Cloud& 17&--12.37&--71.26&7.760E+03&9.568E+03\\
    \hline
     \textbf{ Molli\`ere ``Hybrid" Profile}&\\
     \hline
    Cloudless&17&--36.18&--180.36&7.751E+03&9.505E+03\\
    Grey Deck Cloud&19&--2.13&--60.36&7.785E+03&9.622E+03\\
    Grey Slab Cloud&20&--10.97&--95.00&7.776E+03&9.588E+03\\
    Power Law Deck Cloud&20&0.00&0.00&7.778E+03&9.683E+03\\
    Power Law Slab Cloud&21&--11.81&--18.19&7.775E+03&9.664E+03\\
    \hline
    \end{tabular}
    \caption{List of models and thermal profiles tested in this work for \BD~and \WISE~along with the  corresponding $\Delta$logEv. The lowest $\Delta$logEv (e.g., 0.0) corresponds to the `winning'' model for each respective thermal parameterization for \BD~and \WISE. }
    \label{tab:models_tested}
\end{table*}




\section{\BD~Retrieval Results}
\label{sec:BD60_retrieval_results}

We show the list of models tested for \BD~with the number of parameters and $\Delta$logEv for each model and thermal profile parameterization in Table \ref{tab:models_tested}.

\subsection{Best-fit Models}

The top-ranked model for \BD~was the power law deck cloud model when using either the simple five point spline or the Molli\`ere ``Hybrid" thermal profile.  When using the Lavie thermal profile, the grey deck cloud model was top-ranked and the power law deck cloud was a close third. With $\Delta$logEv greater than 2 in all three cases, it is clear that the cloudless model is strongly rejected, indicating that the data strongly favor a cloud model to fit the spectroscopic features observed in \BD.

\subsection{Thermal Profiles}

\begin{figure}[p]
\gridline{\fig{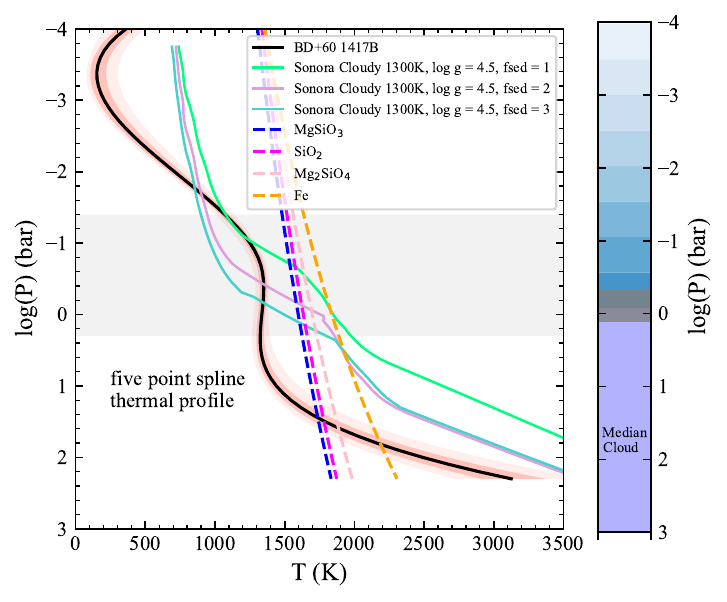}{0.37\textwidth}{\small(a)}}
\gridline{\fig{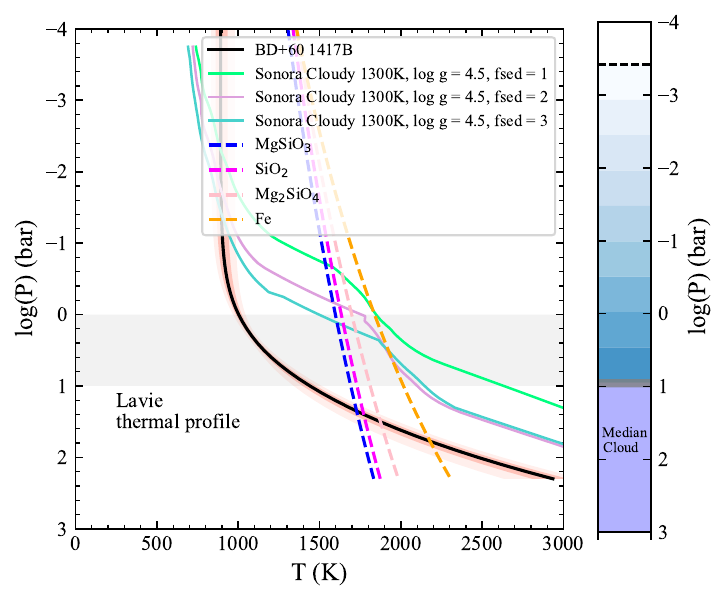}{0.37\textwidth}{\small(b)}}
\gridline{\fig{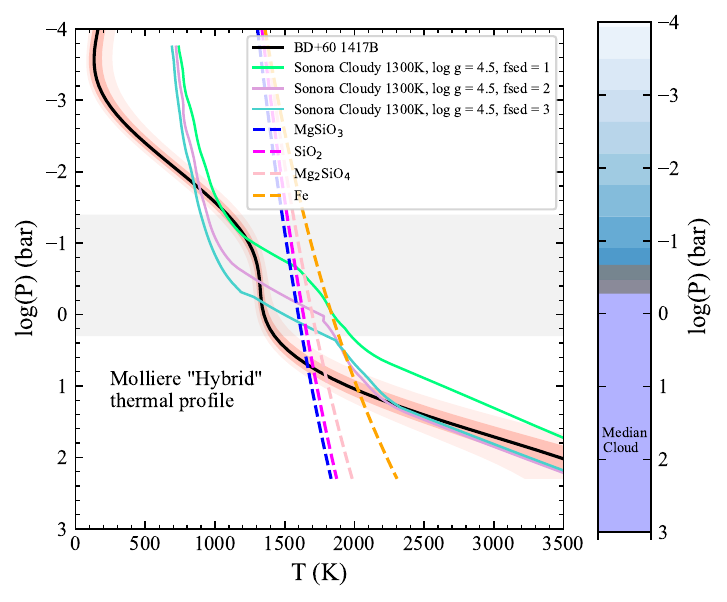}{0.37\textwidth}{\small(c)}}
 \caption{Retrieved thermal profile (black line, light-red shading for 1$\sigma$ and 2$\sigma$ intervals) and cloud pressure for \BD~of the winning models for tested thermal profiles. Dashed lines show the condensation curves for possible cloud species. 
The lightgray region is the approximate photosphere.  Self-consistent grid models from Sonora Diamondback~\citep{Morley2024} for SED derived temperature are plotted as solid colored lines. The median pressures of the extent of the cloud is shown in purple shading with gray shading indicating the  1$\sigma$ range, and the blue shading shows the vertical distribution of the deck cloud over which the optical depth drops to 0.5. (a) \BD~retrieved thermal profiles for five point spline, (b) Lavie profile, (c) Molli\`ere ``Hybrid" profile
\label{fig:thermal_profs_BD60}}
\end{figure}

We show the retrieved thermal profiles and cloud pressures for the winning models compared to  Sonora Diamondback self-consistent grid models~\citep{Morley2024} and condensation curves for \BD~in Figure \ref{fig:thermal_profs_BD60}. For the grid models shown, the base model has a $T_\mathrm{{eff}}$ = 1300 K, log $g$ = 4.5, and a sedimentary efficiency, $f_\mathrm{sed}$ = 1--3. These values are selected from the SED-derived parameters for \BD.  
\par
For the five point spline profile~(Figure \ref{fig:thermal_profs_BD60}a)  there exists a slight thermal inversion near the top of the atmosphere, which we conclude is unphysical. This phenomenon is a byproduct of the five point ``irregular" thermal profile~\citep{Calamari2022}. Additionally, we note that the retrieved thermal profile is not in agreement with the Sonora Diamondback grid models at the predicted temperature, especially within the photosphere~(indicated by light grey shading) where our retrieved profile exhibits primarily warmer temperatures than the models predict. In the bottom half of the observable photosphere, our retrieved profile is nearly isothermal.  Near the top of the visible photosphere region, the $f_\mathrm{sed}$ = 1 model is near the same temperature as our retrieved profile.  The location of the median cloud deck (1$\sigma$) is within the observable photosphere for the five point thermal parameterization. Generally we find that the best fit retrieval using the five point spline thermal profile  for \BD~looks physically implausible (Figure \ref{fig:thermal_profs_BD60}a)


Figure \ref{fig:thermal_profs_BD60}b shows the retrieved thermal profile for the winning model for the Lavie thermal profile. The Lavie profile is a less flexible profile, which forces an almost isothermal profile above the photosphere. In the photosphere region, the thermal profile for \BD~is cooler than Sonora Diamondback grid models predictions. The cloud top is near the bottom of the photosphere.

In Figure \ref{fig:thermal_profs_BD60}c we show the retrieved thermal profile and cloud pressure for the winning model for \BD~for the Molli\`ere thermal parameterization.  In the photosphere region, the retrieved profile is nearly isothermal and the median cloud lies within the observable photosphere, similar to the five-point spline model.  In the regions extrapolated below the observable photosphere, the $f_{\mathrm{sed}}=2$ and $f_{\mathrm{sed}}=3$ models most closely trace the extrapolated profile.  

\subsection{Contribution Functions and Retrieved Spectra:}
\label{sec:retrieved_spectra_BD60}

The contribution function in a layer is defined as
\begin{equation}
\label{eqn:contributionlayers}
    C(\lambda, P) =\frac{B(\lambda,T(P))\int_{P_1}^{P_2}d\tau}{\exp{\int_0^{P_2}d\tau}}
\end{equation}
where $B(\lambda,T(P))$ is the Planck function, zero is the pressure at the top of the atmosphere, $P_1$ is the pressure at the top of the layer, and $P_2$ is the pressure at the bottom of the layer. The contribution function indicates the location in the atmosphere where observable emission features are produced from and where spectral features become muted. The  $\tau$ = 1 gas and cloud contribution for winning models are shown as well. 

\par

\begin{figure}[p]
\gridline{\fig{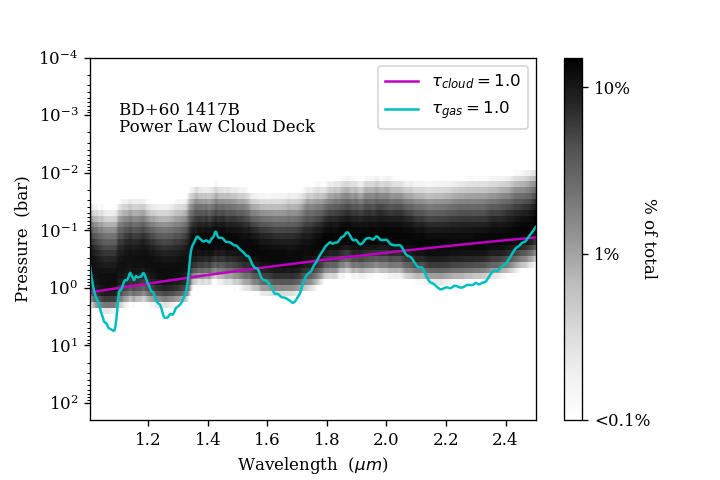}{0.50\textwidth}{\small(a)}}
     \gridline{\fig{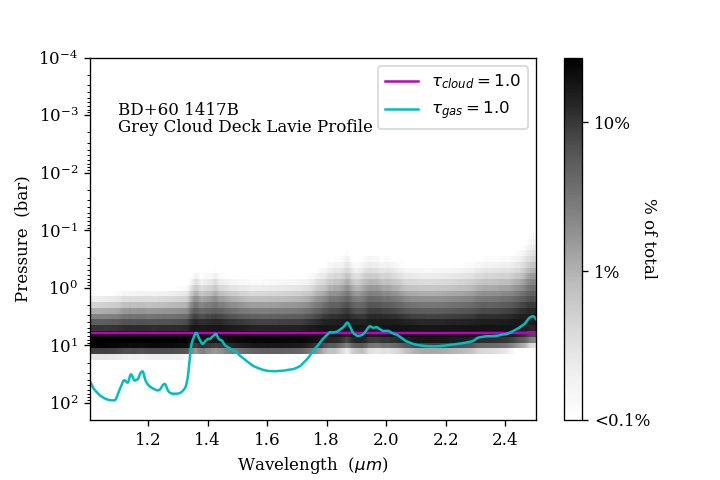}{0.50\textwidth}{\small(b)}}
     \gridline{\fig{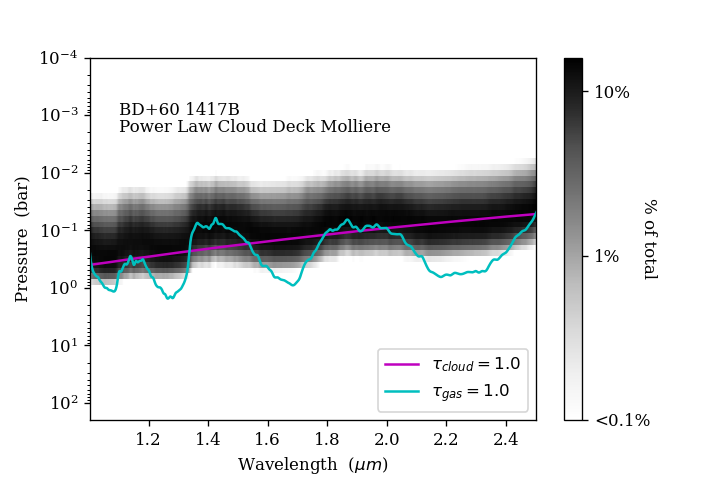}{0.50\textwidth}{\small(c)}}
 \caption{Contribution functions for \BD~based on top-ranked model for each thermal parameterization case. The black shading shows the percentage contribution of each pressure to the flux at each wavelength. $\tau=1$ lines are included for gas phase opacities (cyan),  and a cloud (magenta) (a) \BD~contribution function for five point spline thermal profile, (b) Lavie profile, (c) Molli\`ere ``Hybrid" profile}
   \label{fig:contribution_functions_BD60B}
\end{figure}

Figure \ref{fig:contribution_functions_BD60B}a shows the contribution function for the five point spline thermal profile model for \BD~for the winning model of the power law cloud deck. The $\tau$ = 1 gas and cloud contributions are shown. The majority of the flux contribution to the observed spectrum comes from approximately 0.004 to 2 bar region, corresponding to the photosphere. The $K$ band is shaped by the cloud opacity, which has a more significant contribution. In the $J$ and $H$ bands, the gas opacity contributes more, which is the wavelength regions corresponding to H$_{2}$O molecular band features. There are windows in the $J$ ($\approx$ 1.2 to 1.4 $\mu$m) and $H$ ($\approx$ 1.6 to 1.8 $\mu$m) bands where the cloud opacity dominates.


Figure \ref{fig:contribution_functions_BD60B}b, shows the contribution function for the Lavie thermal profile for \BD~for the winning model, the power law cloud deck. The bulk of the flux between $\approx$ 0.7 and 20 bars contributes to the observed spectrum. Across the spectrum, the cloud opacity becomes more important, with the exception of a small region in the H band ($\approx$ 1.8 -- 2.0 $\mu$m), where the gas opacity dominates.

Figure \ref{fig:contribution_functions_BD60B}c shows the contribution function for the Molli\`ere thermal profile. The contribution function is similar to the five point spline thermal profile. One of the key differences is the dominance of the cloud opacity in the entire $J$ band region. Between 1.3 -- 1.5 $\mu$m is the only region where the gas opacity has a larger contribution.

\begin{figure}[p]
\gridline{\fig{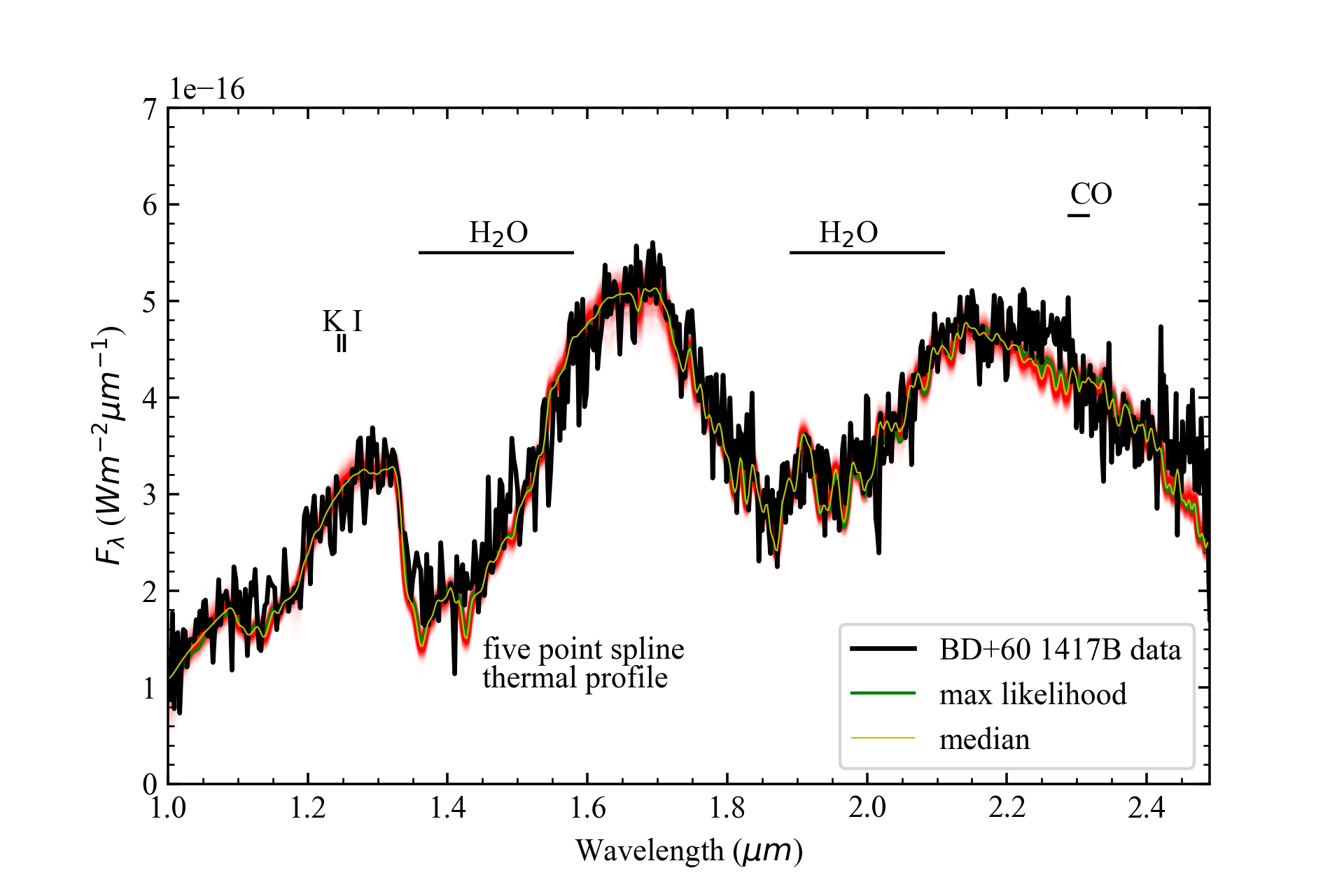}{0.50\textwidth}{\small(a)}}
\gridline{\fig{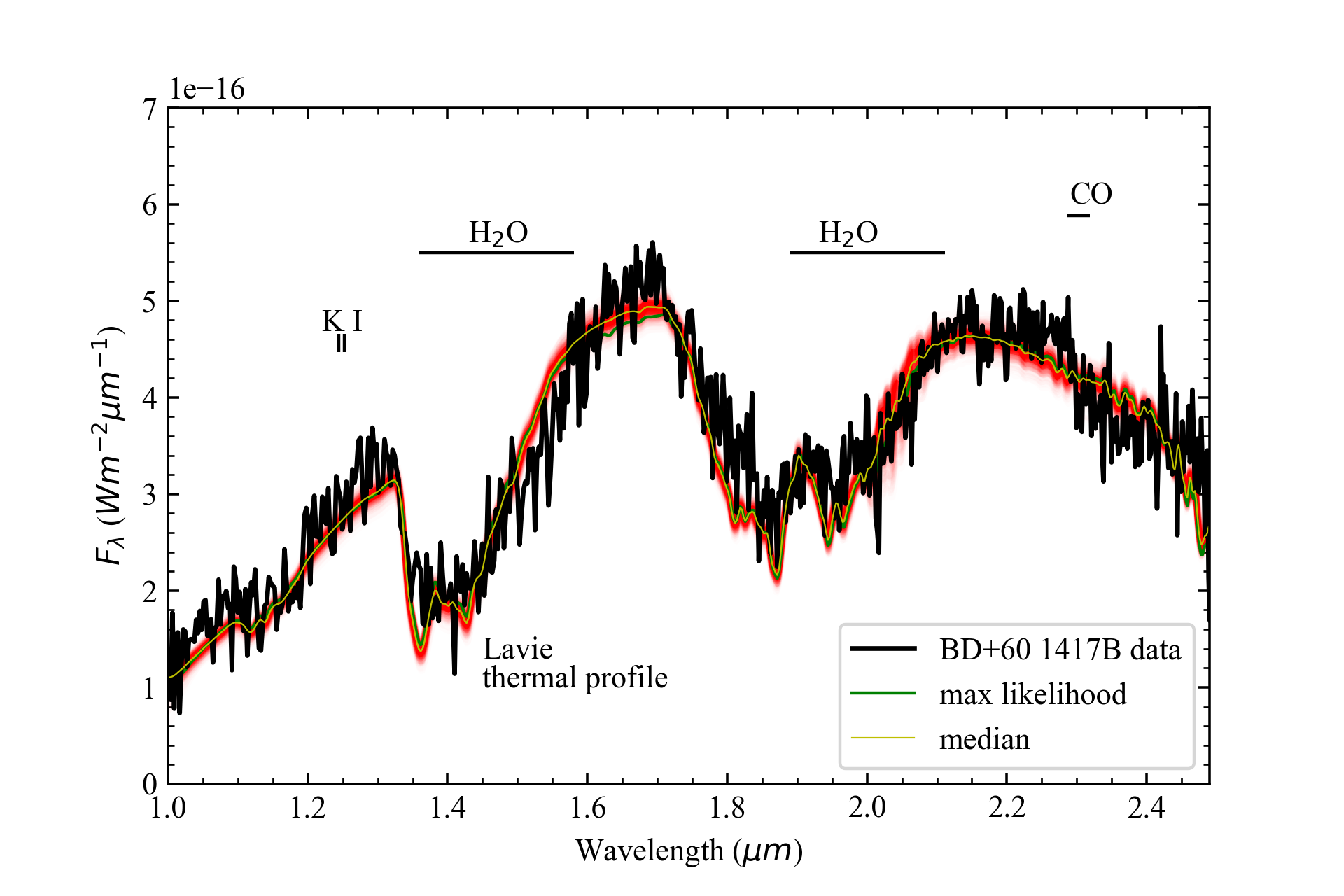}{0.50\textwidth}{\small(b)}}
\gridline{\fig{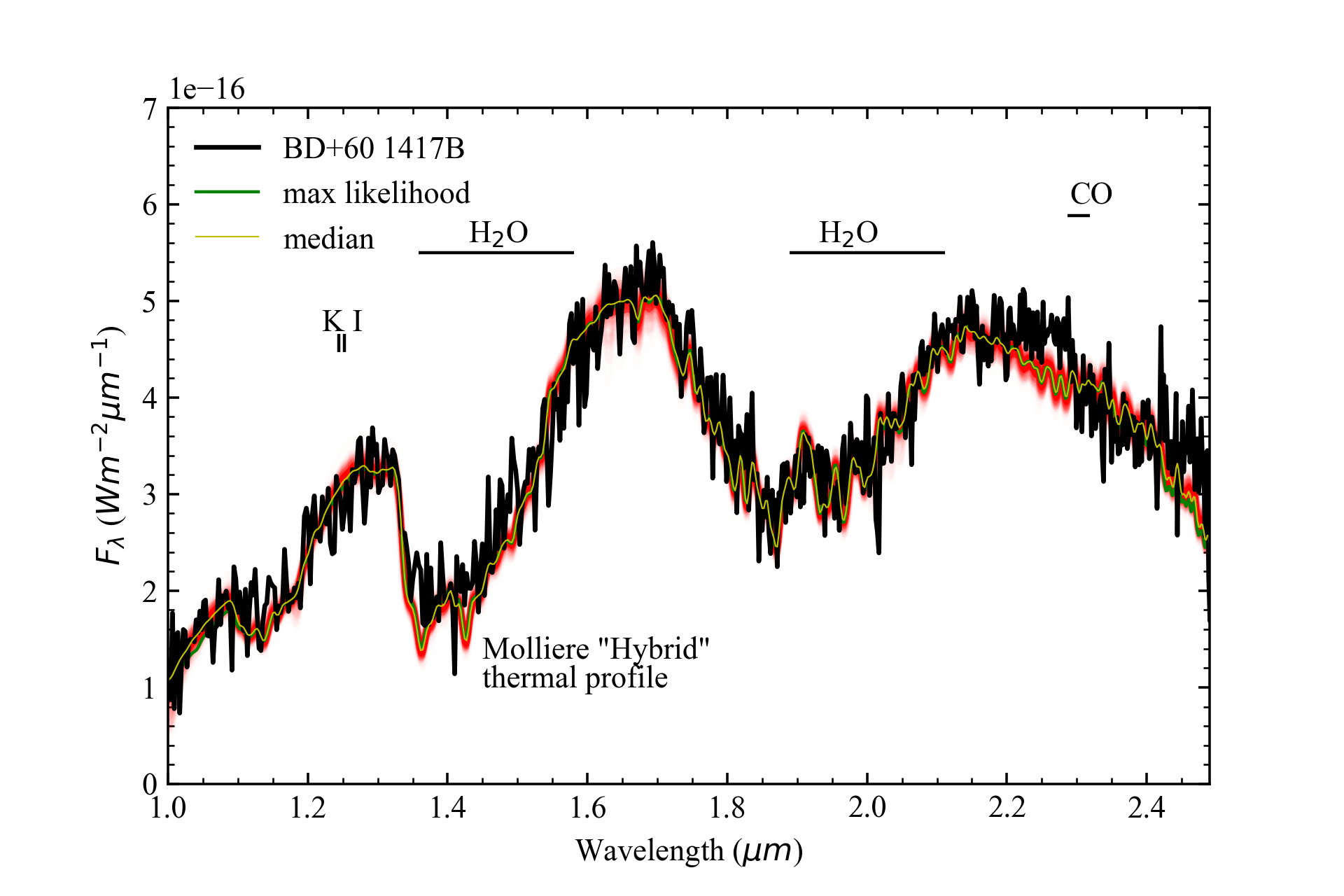}{0.50\textwidth}{\small(c)}}
 \caption{Retrieved forward model spectra for \BD.  The maximum-likelihood spectrum is shown in green and the median spectrum in yellow. The SpeX prism data is shown in black. (a) \BD~retrieved spectra for five point spline thermal profile, (b) Lavie profile, (c) Molli\`ere ``Hybrid" profile. }
   \label{fig:retrieved_spectra_BD60B}
\end{figure}

We show all the retrieved spectra for \BD~in Figure \ref{fig:retrieved_spectra_BD60B} for our winning fits. The retrieved forward model spectra consist of both the maximum-likelihood spectrum and median spectra. The maximum likelihood spectrum is the spectrum that maximizes the likelihood (i.e., best fit from every single iteration). In  the retrieval, the median spectrum is the spectrum from the median value of each retrieved parameter which is put into the forward model. Utilizing both the maximum-likelihood and the median spectra explores the parameter space well and ideally these two retrieved fits should be near identical.
\par
Figure \ref{fig:retrieved_spectra_BD60B}a shows the retrieved spectrum for \BD~for the five point spline thermal profile. The retrieved spectrum shows good agreement through the $J$ and $H$ spectral bands. However, the retrieved spectrum shows a worse fit in the $K$-band which holds the temperature dependent CO molecular bands. The retrieved spectrum appears too faint for the $K$-band overall, which may explain our poorly constrained CO abundance in the five point spline thermal parameterization.

Figure \ref{fig:retrieved_spectra_BD60B}b shows the retrieved spectrum for the Lavie profile for \BD. We note that the retrieved spectrum  is fainter than the flux peaks in the $H$ spectral band which contains the gravity sensitive triangular-band shape, this is in contrast to the five point spline thermal profile. Similarly, the $K$ spectral band is poorly fit as well.

Figure \ref{fig:retrieved_spectra_BD60B}c shows the retrieved spectrum for the Molli\`ere thermal profile for \BD. Similar to the Lavie retrieved profile~(Figure \ref{fig:retrieved_spectra_BD60B}b) the $K$ band is too faint, which contains the CO band head features. 


\subsection{Retrieved Gas Abundances and Fundamental Parameters:}

Figure \ref{fig:corner_BD60_comparison} shows the posterior distributions for the radius, mass, metallicity, C/O,  and $T_{\rm eff}$ for the retrieval case of \BD. 
The derived log(g) is calculated from the retrieved radius and mass. The scale factor ($R^2/D^2$) is calculated from the retrieved radius and parallax. $T_{\rm eff}$ is then determined using this derived scale factor and by integrating the flux in the resultant forward model spectra extrapolated to cover 0.5 -- 20 $\mu$m. 

Determining the carbon-to-oxygen ratios (hereafter C/O) of substellar objects involves retrieving the following abundances ({$X$}):

\begin{equation}
\label{eq:CtoO}
    \cfrac{\rm C}{\rm O} = \cfrac{X_{\rm CO}+X_{\rm CO_{2}}+X_{\rm CH_{4}}}{X_{\rm CO}+2X_{\rm CO_{2}}+X_{\rm H_{2}O}}
\end{equation}

Given our limited spectral coverage and retrieved abundances, when appropriate the C/O ratio is calculated under the assumption that all of the oxygen exists in H$_2$O and CO and all of the carbon exists in CO.  Our ratio is considered as a CO to H$_{2}$O ratio. We note however that broader wavelength range retrievals for L dwarfs (e.g. \cite{Burningham2021} suggest that CO$_{2}$ abundances are higher than expected and may need to be taken into account for this total ratio. 

\begin{equation}
\label{eq:CtoO}
    \cfrac{\rm C}{\rm O} = \cfrac{X_{\rm CO}}{X_{\rm CO}+X_{\rm H_{2}O}}
\end{equation}
\par
The following equations were used to derived a value for metallicity, [M/H]:
\begin{subequations}\label{eq:metallicity}
\begin{equation}
    \mathit{f}_{\rm H_2} = 0.84(1 - \mathit{f}_{gas})
\end{equation}
\begin{equation}
    N_H = 2\mathit{f}_{\rm H_2}N_{tot}
\end{equation}
\begin{equation}
    N_{\rm element} = \sum_\mathrm{{molecules}} n_\mathrm{{atom}}\mathit{f}_\mathrm{{molecule}}N_\mathrm{{tot}}
\end{equation}
\begin{equation}
    N_{\rm M} = \sum_\mathrm{{element}} \frac{N_\mathrm{{element}}}{N_{\rm H}}
\end{equation}

\end{subequations}
where $\mathit{f}_{H_2}$ is the $H_2$ fraction, $\mathit{f}_{gas}$ is the total gas fraction for all other gases, 0.84 represents the fraction of gas that is H$_{2}$, $N_H$ is the number of neutral hydrogen atoms, $N_{element}$ is the number of atoms for each element of interest, $n_{atom}$ is the number of atoms for a given element contained in a single molecule and $N_{tot}$ is the total number of gas molecules.
\par
In our calculation of metallicity, $N_{solar}$ is determined using the same formula as $N_{\rm 
 M}$, using the sum of the solar abundances (from \citealt{Asplund2009}) relative to H. 

\begin{equation}
    [{\rm M/H}] = \log\frac{N_{\rm M}}{N_{\odot}}
\end{equation}
\begin{deluxetable}{lrr}[tb]
\tabletypesize{\small}
\tablecolumns{10}
\tablewidth{0pt}
\setlength{\tabcolsep}{0.05in}
\tablecaption{Comparison off retrieved gas abundances for the preferred models for \BD~and \WISE~for our various thermal profile parameterizations}
\label{tab:retrieved_gas_abundances}
\tablehead{  
\colhead{}&
\colhead{\BD}&
\colhead{\WISE}}
\startdata
\colhead{\textbf{5 point Profile}}\\
\hline
H$_2$O      & $-0.40^{+0.21}_{-0.27}$         & $-0.68^{+0.22}_{-0.20}$          \\
CO          & $-7.38^{+3.00}_{-2.83}$         & $-8.96^{+2.44}_{-1.69}$          \\
CO$_2$      & $-5.73 ^{+2.60}_{-3.13}$         & $-6.68^{+2.32}_{-2.74}$          \\
CH$_4$      & $-8.16^{+2.44}_{-2.37}$         & $-7.60^{+3.35}_{-2.34}$          \\
CrH         & $-8.05^{+2.44}_{-2.37}$         & $-5.90^{+0.27}_{-0.31}$          \\
FeH         & $-9.07^{+1.98}_{-1.96}$         & $-8.67^{+1.34}_{-1.70}$          \\
Na+K        & $-7.90^{+3.30}_{-2.64}$         & $-8.45^{+2.46}_{-1.92}$          \\ 
\hline
\colhead{\textbf{Lavie Profile}}\\
\hline
H$_2$O      & $-3.34^{+0.05}_{-0.05}$         & $-1.48^{+0.12}_{-0.06}$          \\
CO          & $-8.15^{+2.46}_{-2.10}$         & $-1.76^{+0.21}_{-0.39}$          \\
CO$_2$      & $-5.78  ^{+2.46}_{-3.97}$         & $-4.13^{+0.78}_{-0.97}$          \\
CH$_4$      & $-8.31^{+1.78}_{-1.96}$         & $-3.10^{+0.19}_{-0.23}$          \\
CrH         & $-9.39 ^{+1.00}_{-1.39}$         &  $-10.94 ^{+0.48}_{-0.35}$          \\
FeH         & $-10.07^{+0.85}_{-0.98}$         & $-10.56^{+0.67}_{-0.50}$          \\
Na+K        & $-8.58 ^{+2.15}_{-1.94}$         & $-2.32^{+0.18}_{-0.15}$          \\ 
\hline
\colhead{\textbf{ Molli\`ere ``Hybrid" Profile}}\\
\hline
H$_2$O      & $-0.44^{+0.23}_{-0.29}$         & $-0.58^{+0.24}_{-0.38}$ \\
CO          & $-6.75^{+2.96}_{-2.99}$         & $-1.78^{+0.58}_{-5.55}$          \\
CO$_2$      & $-6.60 ^{+3.26}_{-3.01}$         & $-7.64^{+2.39}_{-2.23}$          \\
CH$_4$      & $-8.07^{+2.30}_{-2.16}$         & $-8.32^{+2.59}_{-2.07}$          \\
CrH         & $-7.33 ^{+1.80}_{-2.57}$         &  $-8.58 ^{+2.47}_{-2.38}$          \\
FeH         & $-8.27^{+1.50}_{-1.91}$         & $-6.69^{+0.43}_{-0.80}$          \\
Na+K        & $-6.69 ^{+3.19}_{-3.05}$         & $-8.59^{+3.00}_{-1.95}$          \\ 
\hline
\enddata
\end{deluxetable}
\par

In Table \ref{tab:retrieved_gas_abundances}, we show the retrieved gas abundances for the winning models for the three thermal profile parameterizations for \BD. 

\begin{figure}[p]
\gridline{\fig{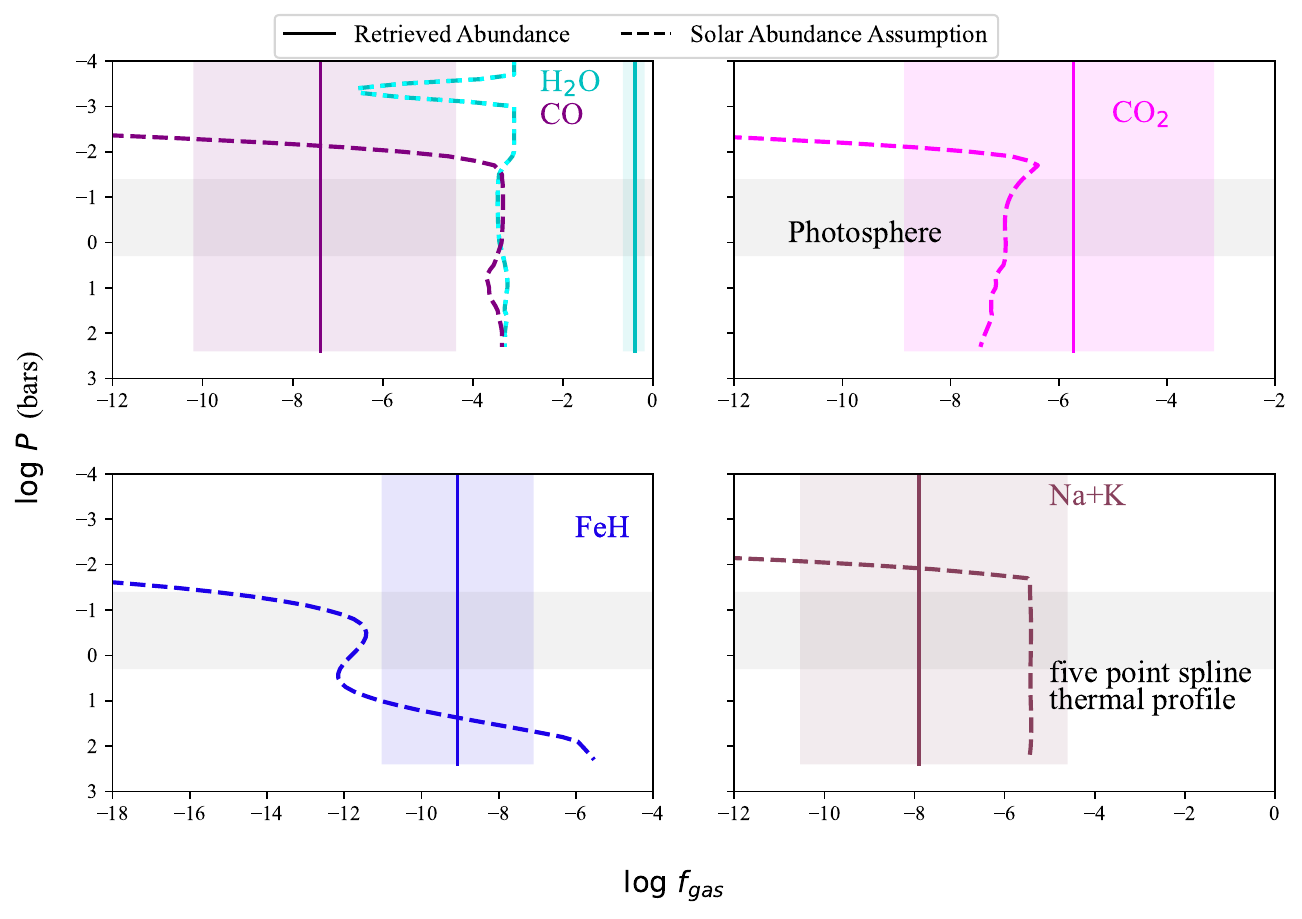}{0.45\textwidth}{\small(a)}}
\gridline{\fig{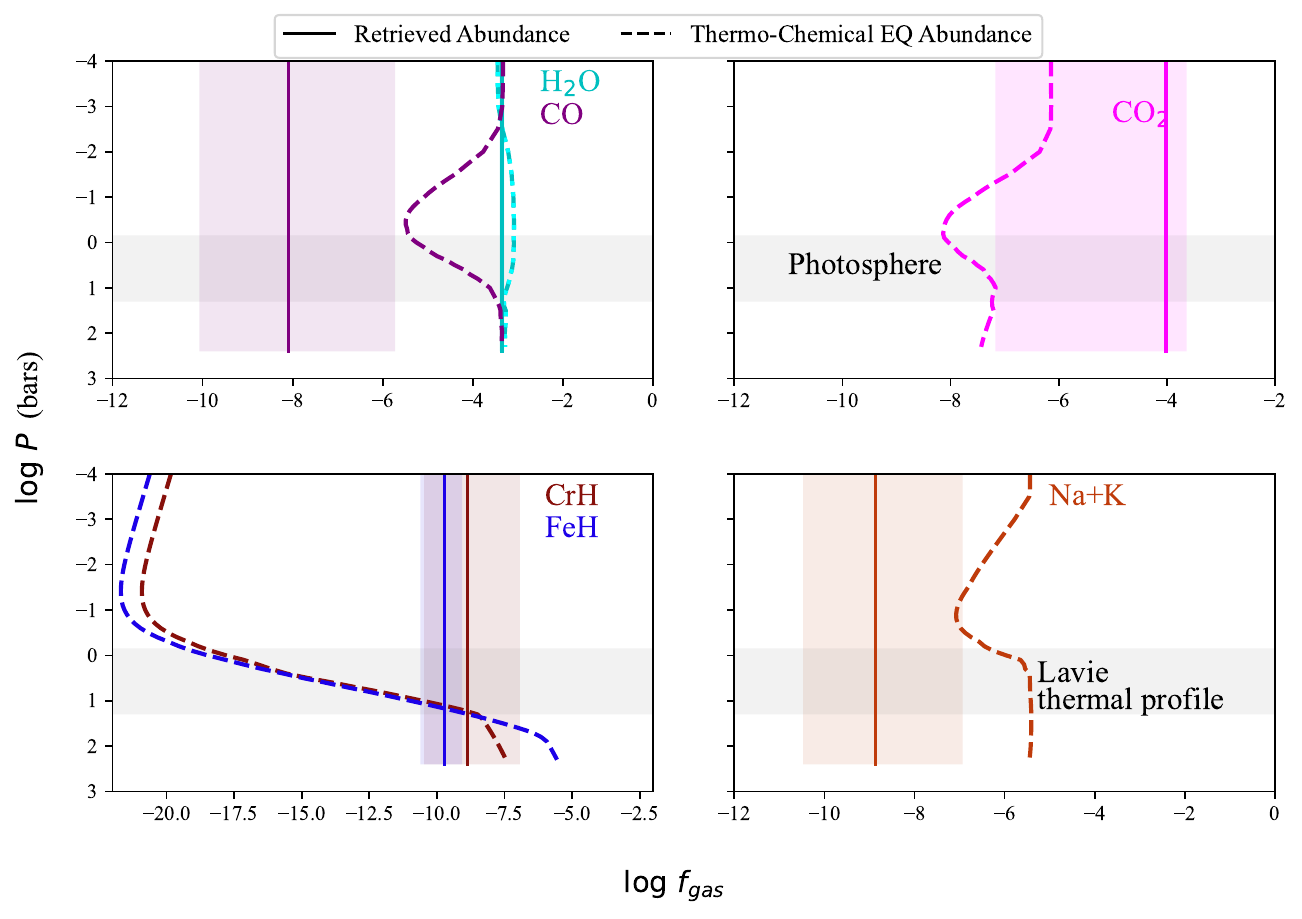}{0.45\textwidth}{\small(b)}}
\gridline{\fig{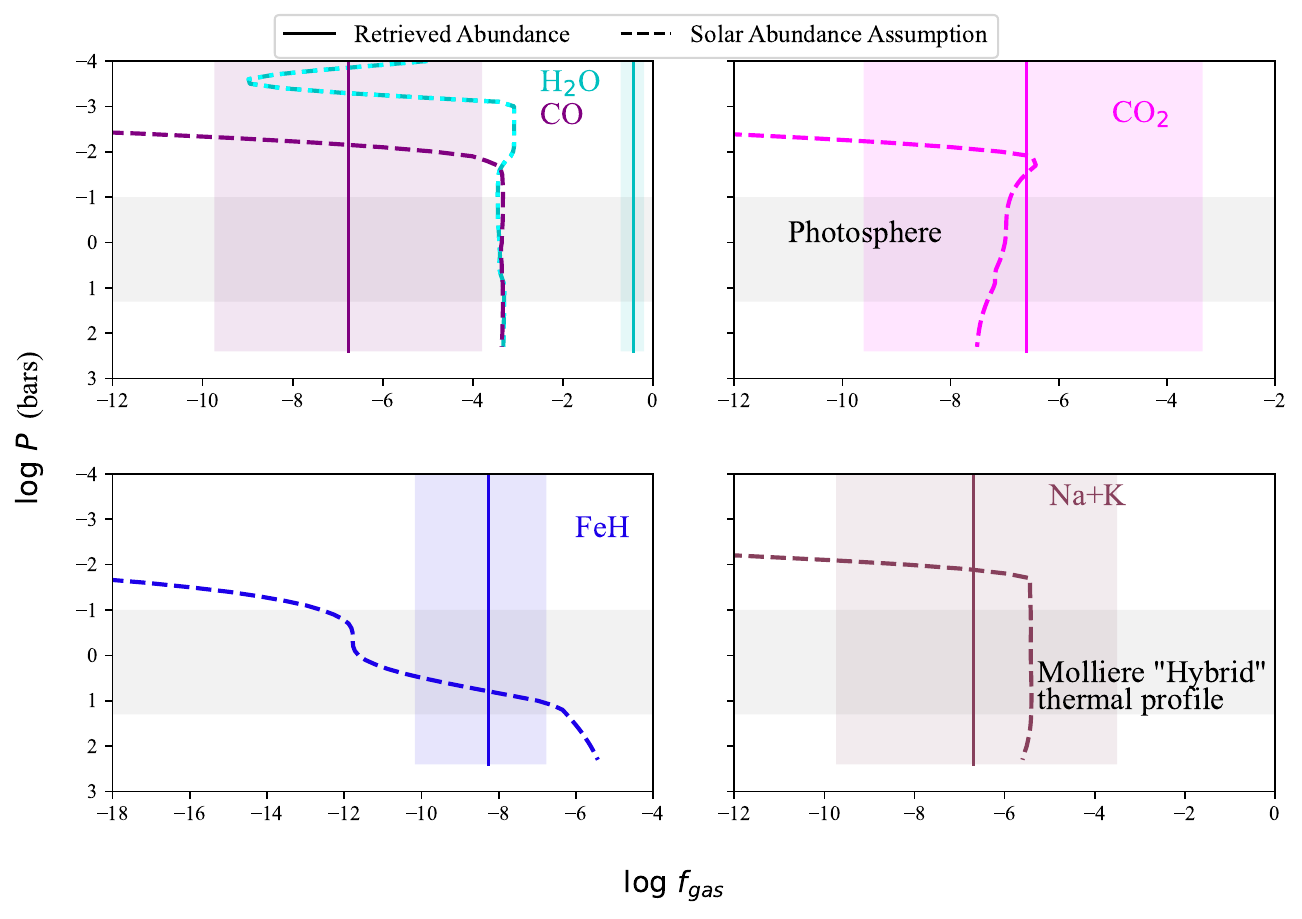}{0.45\textwidth}{\small(c)}}
 \caption{Retrieved uniform-with-altitude mixing abundances (solid lines) compared to thermochemical equilibrium grid abundances (dashed lines) for \BD. Equilibrium predictions are calculated for assumed solar abundance or calculated based on derived [M/H] and C/O values. Shaded regions around the retrieved abundance values show our median retrieved values and 16th to 84th percentiles, respectively. (a) mixing abundances for the five point spline profile, (b) Lavie profile, (c) Molli\`ere ``Hybrid" profile.}
   \label{fig:mix_abundances_BD60B}
\end{figure}

Figure \ref{fig:mix_abundances_BD60B} shows the retrieved gas mixing ratios compared to predictions from thermochemical equilibrium models. The thermochemical equilibrium models are calculated and interpolated for our derived metallicity and C/O ratio for the three thermal profile parameterizations for \BD~where applicable. In the cases where the C/O ratio and [M/H] are unable to be determined (due to nonphysical results) we assume solar values~\citep{Asplund2009} -- this is the case for the five point spline profile and the Molli\`ere ``Hybrid" profile. 
\par
In Figure \ref{fig:mix_abundances_BD60B}a, for the five point spline thermal profile major atmospheric constituents (CO and H$_{2}$O) in the photosphere denoted by the gray box, we see that the gases are not predicted to vary greatly with altitude for our assumed solar abundances. However, our retrieved abundance for H$_{2}$O  is implausibly high and does not match thermochemical equilibrium predictions. For the FeH abundance, our uniform with altitude abundance is in disagreement with the model thermochemical abundance predictions, which is predicted to vary with altitude in the photospheric region.  Conversely, we see that while the Na+K abundance is uniform with altitude in the atmospheres, our median retrieved abundance is depleted compared to the equilibrium predictions. Additionally, we see that our CO measurement for the five point spline thermal profile shows a depletion compared to predictions. This depletion could be the result of poor fits to the $K$ band region of the spectrum for \BD~as shown in Figure \ref{fig:retrieved_spectra_BD60B}a.

\par

In Figure \ref{fig:mix_abundances_BD60B}b we show our retrieved gas mixing ratios along with predictions from thermochemical equilibrium models interpolated for our derived metallicity and C/O ratio for the Lavie thermal profile. For H$_{2}$O, we see a slight disagreement ($>$ 1$\sigma$) in the photospheric region for our retrieved abundance in comparison to the thermochemical equilibrium models. Similar to the five point spline profile, we observe a depletion of predicted CO in the photosphere region that might be due to the poor fits in the $K$ band region. The CO is also predicted to vary with altitude in the photosphere, which cannot be accounted for in our uniform-with-altitude abundance retrieval model. The retrieved FeH abundance shows photospheric agreement near the 10 bar region. While an overall poor constraint, for the CO$_{2}$ retrieved abundances we see an agreement with the predicted abundance that is within the 1$\sigma$ error bar. The Na+K retrieved abundances are depleted compared to thermochemical equilibrium models.
\par

For the Molli\`ere thermal profile, similar to the five point spine thermal profile, the major atmospheric constituents (CO and H$_{2}$O) in the photosphere, we see that the gases are not predicted to vary greatly with altitude for our assumed solar abundances. We again see an implausibly high retrieved H$_{2}$O abundance. For FeH, our retrieved abundance is consistent with the predicted abundance at the bottom of the photosphere at 10 bar. The Na+K retrieved abundance is not predicted to vary with altitude and the values for the thermochemical equilibrium models within the 1$\sigma$ ranges for our retrieved values

 Table \ref{tab:fund_derived_para_compar} shows the comparison of the fundamental and derived parameters for \BD~and the analog object \WISE~for the three different thermal profile parameterizations. We find that for all three of our thermal parameterization checks for \BD~our fundamental parameters are in disagreement with the SED parameters. We discuss this in more detail in Section \ref{sec:discuss_sed_comparision}.


\begin{table*}[h!]
    \centering
    \begin{tabular}{l|cccccc}
    \hline \hline
     \textbf{\BD}&\\
    \hline

    Thermal Profiles &5 point P-T Profile&Lavie Profile&Molli\`ere ``Hybrid" Profile\\
    Preferred Cloud Type&Power Law Deck Cloud&Grey Deck Cloud&Power Law Deck Cloud\\
    
    \hline
    {PdeckC1P1} (bar)&--0.08$^{+0.19}_{-0.25}$&0.95$^{+0.06}_{-0.05}$&--0.46$^{+0.19}_{-0.21}$\\
    Height/$d\log P$ (bar)&4.39$^{+1.49}_{-1.94}$&4.37$^{+1.39}_{-1.56}$&3.91$^{+1.59}_{-1.60}$\\
    PowC1P1 ($\alpha$) &2.86$^{+0.33}_{-0.37}$&n/a&2.58$^{+0.29}_{-0.28}$\\
    \hline
    \textbf{\WISE}&Power Law Deck Cloud&Power Law Deck Cloud&Power Law Deck Cloud\\
    \hline
    PdeckC1P1 (bar)&--0.17$^{+0.07}_{-0.08}$&0.09$^{+0.05}_{-0.08}$&--0.23$^{+0.07}_{-0.08}$\\
     Height/$d\log P$ (bar)&3.34$^{+1.58}_{-1.20}$&4.97$^{+0.55}_{-0.42}$&4.85$^{+1.12}_{-1.46}$\\
    PowC1P1 ($\alpha$)&2.43$^{+0.14}_{-0.14}$&1.55$^{+0.05}_{-0.07}$&2.30$^{+0.14}_{-0.15}$\\
    \hline
    \end{tabular}
    \caption{Summary of retrieved cloud properties for the preferred models of respective thermal parameterization for \BD~and~\WISE. }
    \label{tab:cloud_properties}
\end{table*}

\subsection{Cloud Properties}

We find that the spectrum for \BD~is best described by a cloudy model regardless of thermal profile parameterization. This remains our strongest retrieval conclusion in the absence of being able to definitively converge on either a P-T profile or reasonable chemical abundances. A summary of cloud parameters for \BD~is shown in Table \ref{tab:cloud_properties}. We note a positive slope in the contribution function for the five point spline thermal profile which can be indicative of large particles (1 $\mu$m) in the atmosphere from Mie theory~(e.g. \citealt{Lacyy2020}). Where appropriate the $\alpha$ parameter is positive for the power law. 
\par

The pressure of the cloud in the atmosphere of \BD~is shown in various shaded regions to the right of Figure \ref{fig:thermal_profs_BD60}. The notable difference is that the pressure at which $\tau$ $>$ 1 for the Lavie profile is higher than those for the five point spline and  Molli\`ere ``Hybrid" thermal profile. For \BD~the cloud height is similar for the five point spline and Lavie profile, while the Molli\`ere ``Hybrid" Profile cloud location extends further in the photosphere.

\section{\WISE~Retrieval Results: A Comparative Example}
\label{sec:W0047_retrievals}

The original aim of this work was to use retrieval analysis and the \textit{Brewster} framework to understand the newly discovered young low surface gravity planetary-mass companion \BD. However, we found that despite testing three different thermal parameterizations from the literature, often times not only did our fundamental parameters not match the SED analysis but we arrived at implausibly high or unconstrained molecular abundances for this benchmark object.
\par

In \cite{Faherty2021}, the isolated L-dwarf \WISE~was found to be a close match in spectral morphology and fundamental characteristics to \BD~(Table \ref{tab:props}). \WISE~is an isolated object with higher S/N SpeX prism observations (S/N $\sim$ 5 vs S/N $\sim$ 13) than \BD, and slightly brighter apparent magnitude. We aim to use the same retrieval framework on this isolated, clone object to see if more robust results are obtained on higher quality data. Table \ref{tab:models_tested} shows the list of models tested for \WISE, along with the number of parameters for each model and the resulting $\Delta$logEv. 
\subsection{Best-fit Models}

The power law deck cloud is preferred by all three thermal parametrizations tested. For this comparative object to \BD~it is again evident that based on our selection criterion, cloudless models are strongly rejected for all three different thermal parameterizations. Once again this emphasizes our strongest retrieval conclusions: \WISE~ is a cloudy L dwarf. \WISE~is known to be variable in the infrared, likely driven by non-uniform condensate clouds~(\citealt{Lew2016}; \citealt{Vos2018}) thus our retrieval conclusion is not unexpected.
\subsection{Thermal Profiles}

\begin{figure}[H]
\gridline{\fig{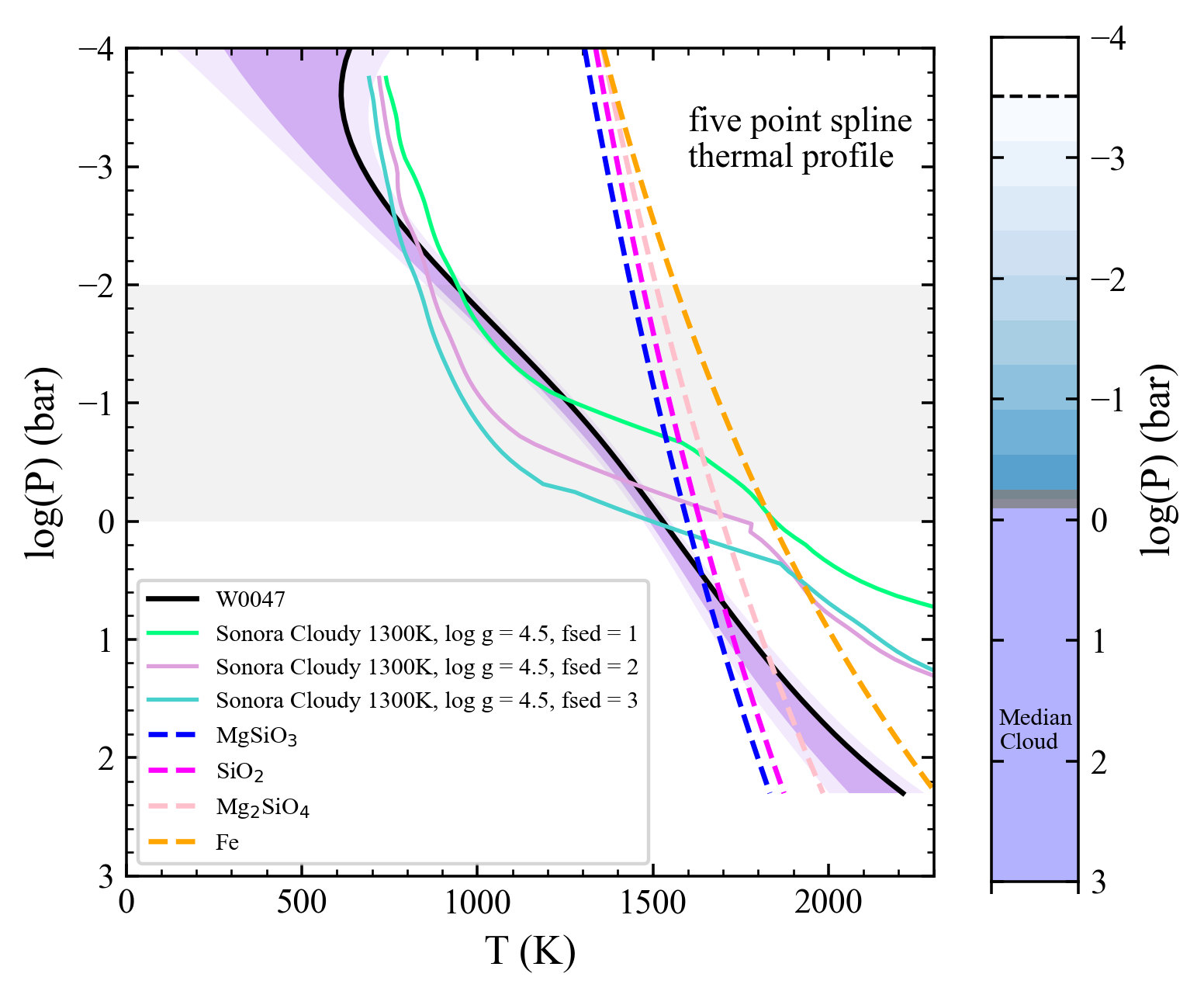}{0.37\textwidth}{\small(a)}}
\gridline{\fig{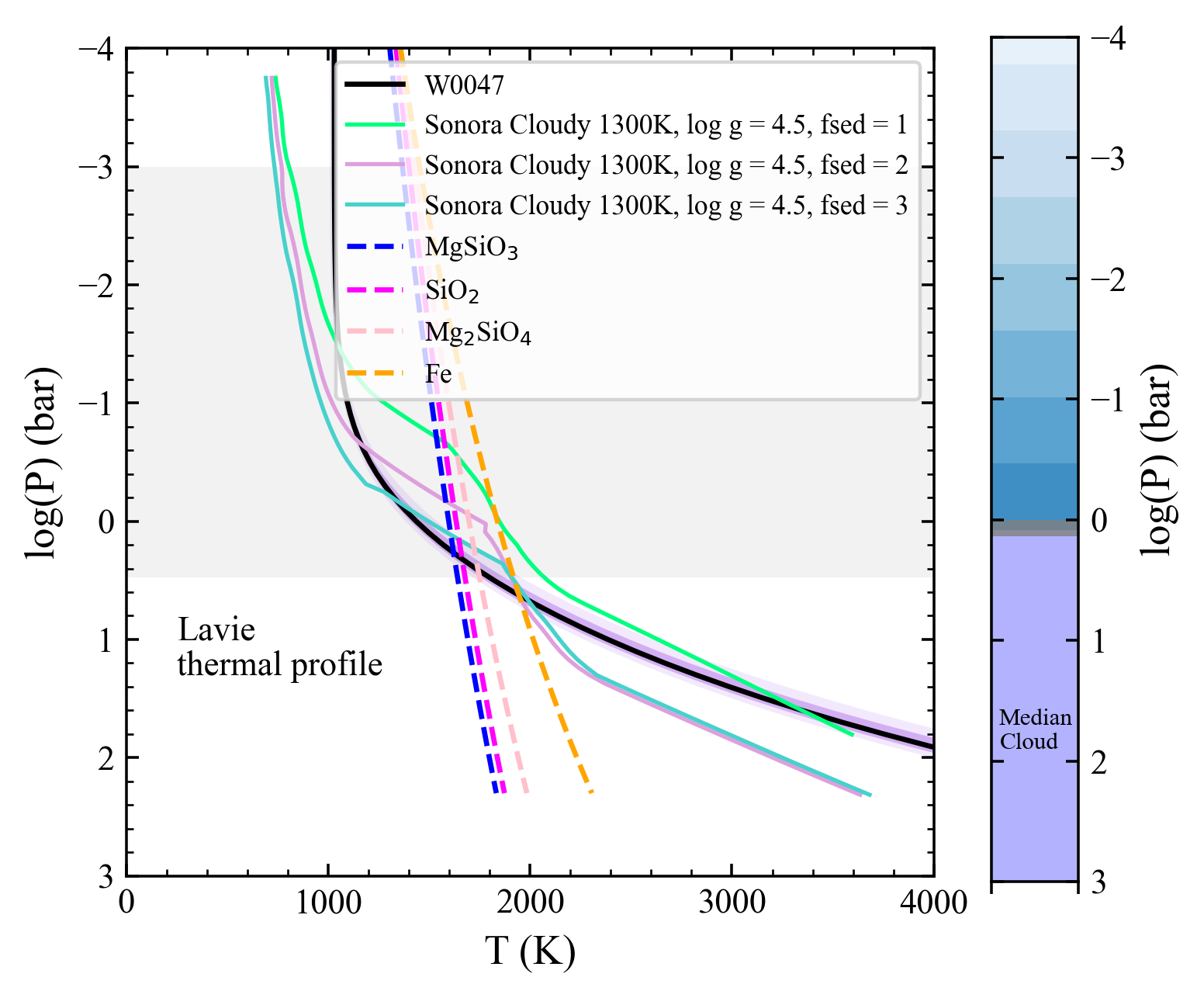}{0.37\textwidth}{\small(b)}}
\gridline{\fig{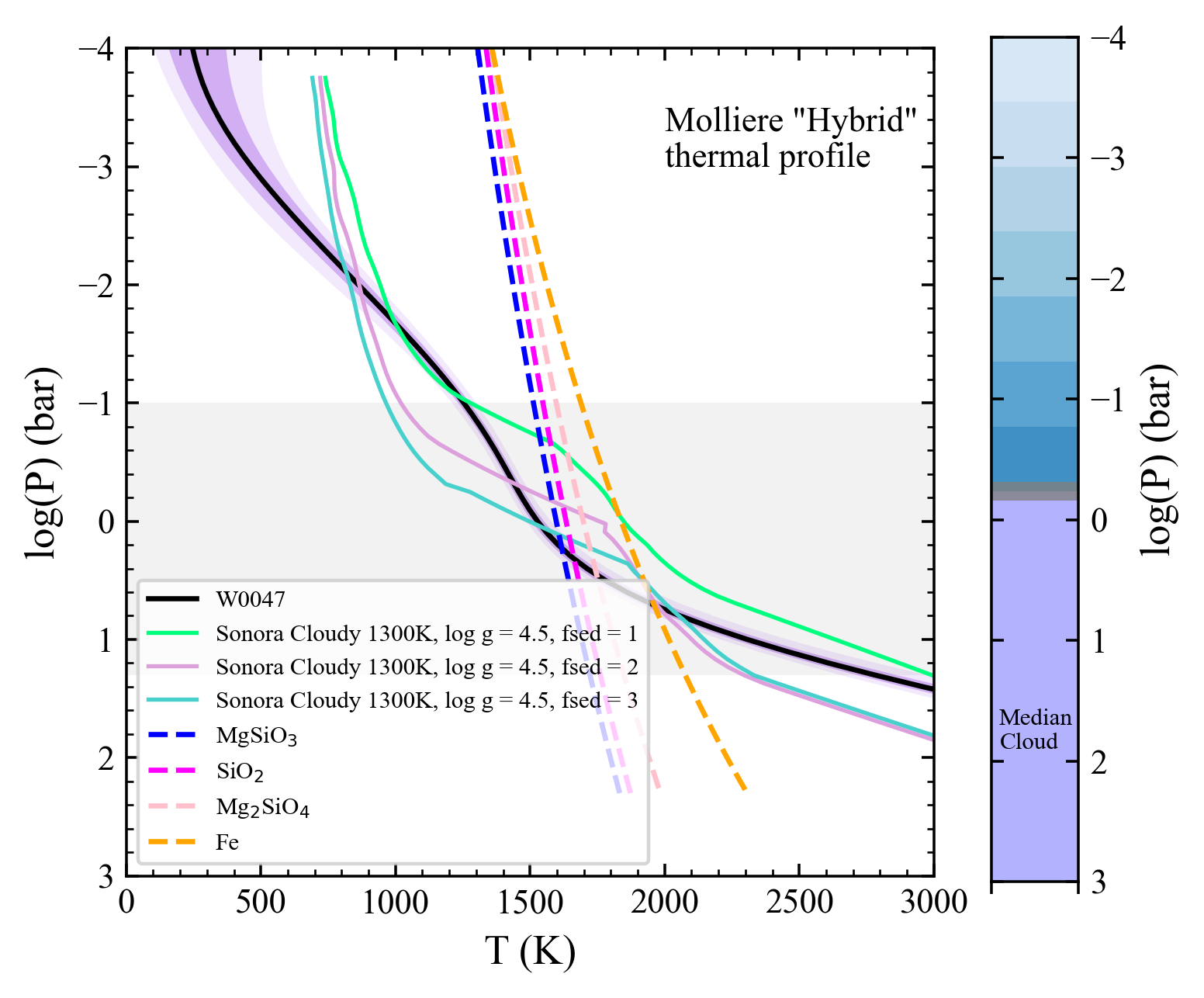}{0.37\textwidth}{\small(c)}}
 \caption{Retrieved thermal profile (black line, light-purple shading for 1$\sigma$ and 2$\sigma$ intervals) and cloud pressure for \WISE~for a power law cloud deck for the Lavie thermal profile. Dashed lines show the condensation curves for possible cloud species. The lightgray region is the approximate photosphere. Self-consistent grid models from Sonora Diamondback profiles for the SED derived temperature are shown as solid colored lines~\citep{Morley2024}. The median pressures of the extent of the cloud is shown in purple shading with gray shading indicating the  1$\sigma$ range, and the blue shading shows the vertical distribution of the deck cloud over which the optical depth drops to 0.5. (a) \WISE~thermal profiles for five point spline, (b) Lavie profile, (c) Molli\`ere ``Hybrid" profile.}
\label{fig:thermal_profs_W0047}
\end{figure}

Figure \ref{fig:thermal_profs_W0047} shows the retrieved thermal profiles and cloud pressures for the winning models compared to grid models and condensation curves for \WISE. For the five point spline (Figure \ref{fig:thermal_profs_W0047}a),  the retrieved thermal profile is not in agreement with the grid models, as was seen for \BD.  In the photosphere, the retrieved thermal profile is $\sim$ 300 -- 400K warmer than the grid models. The shape of the retrieved thermal profile is nearly linear throughout the photosphere and the extrapolated regions above and below the photosphere.  The grid model that has the closest fit to retrieved thermal profile in the photosphere is the $f_{\mathrm{sed}} = 1$ model. The median cloud location 1$\sigma$ range is within the photosphere. 
\par
For the Lavie thermal profile parameterization~(Figure \ref{fig:thermal_profs_W0047}b), we find that the retrieved thermal profile is not well aligned with the grid models. The cloud is located within the observable photosphere. Below the observable photosphere, the closest fitting model is $f_{\mathrm{sed}} = 1$. 
\par


Figure \ref{fig:thermal_profs_W0047}c shows the retrieved thermal profile for the Molli\`ere thermal parameterization. The models ($f_{\mathrm{sed}} = 1$ -- 3) are generally not in agreement with the retrieved profile in the photosphere region. Above the photosphere in the extrapolated region, the models diverge more from the extrapolated thermal profile. For the Molli\`ere profile, the cloud deck is prevalent within the photosphere.

\subsection{Contribution Functions and Retrieved Spectra}
\label{sec:retrieved_spectra_W0047}

\begin{figure}[H]
    \centering
\gridline{\fig{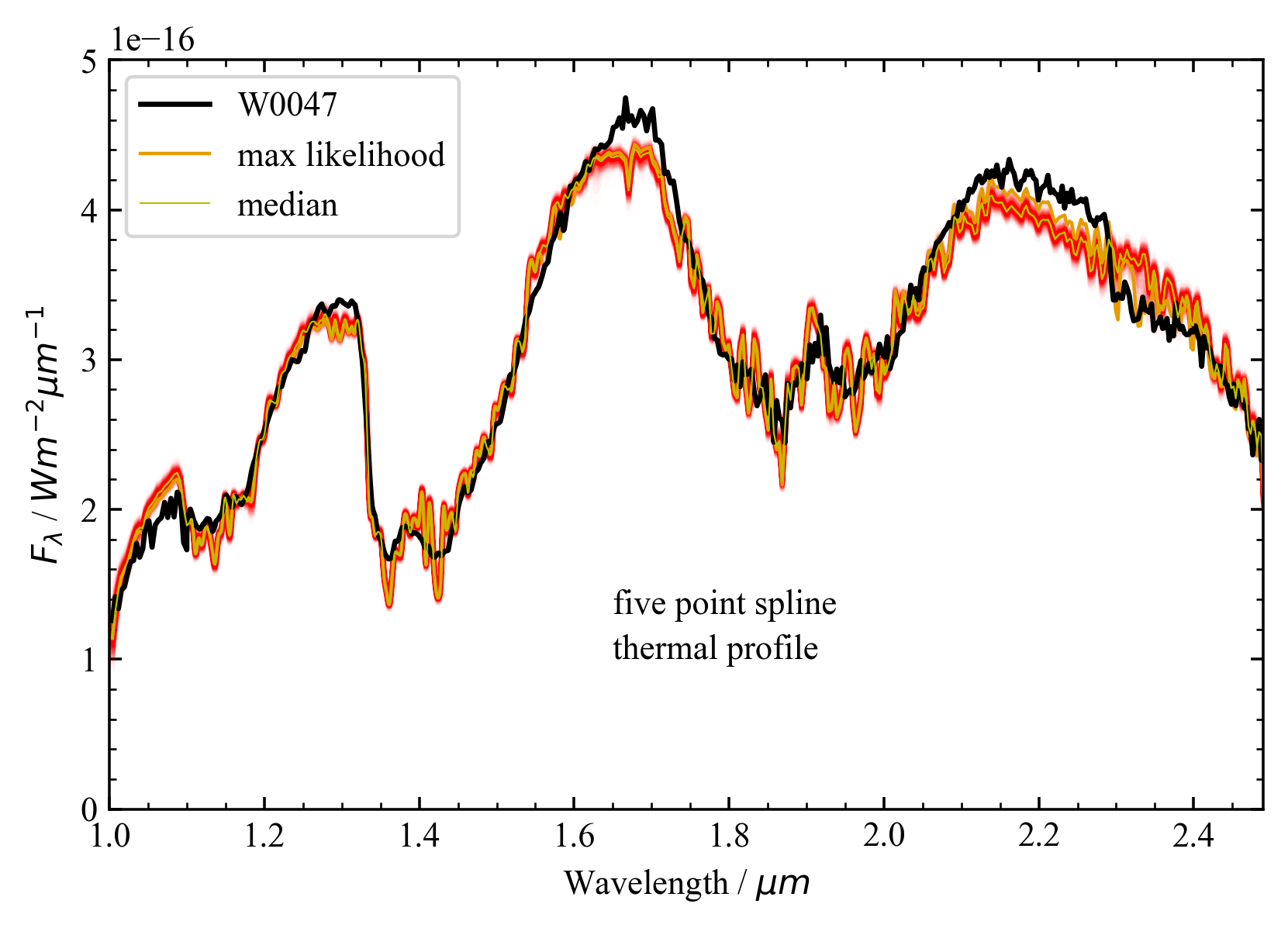}{0.45\textwidth}{\small(a)}}
    \gridline{\fig{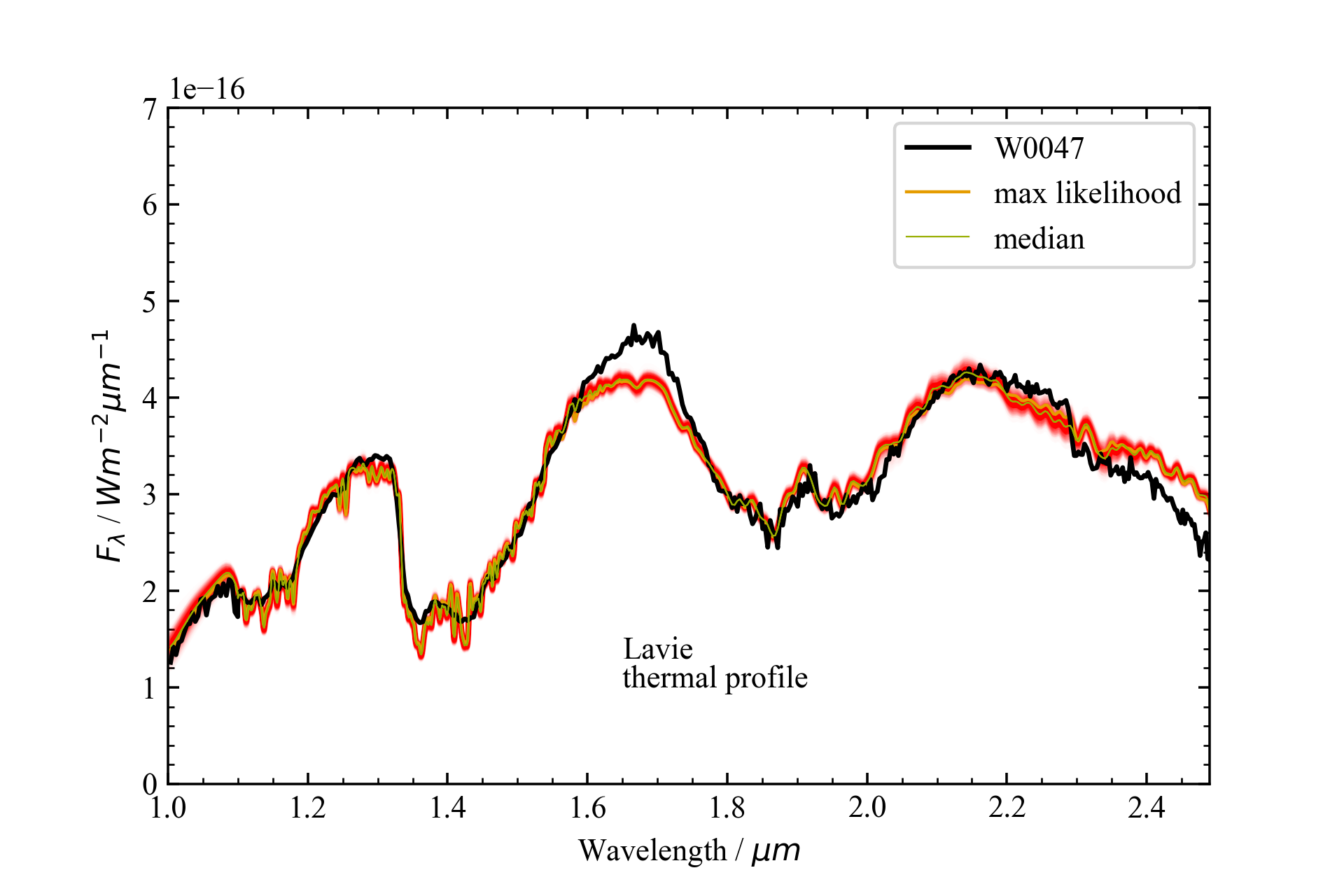}{0.50\textwidth}{\small(b)}}
    \gridline{\fig{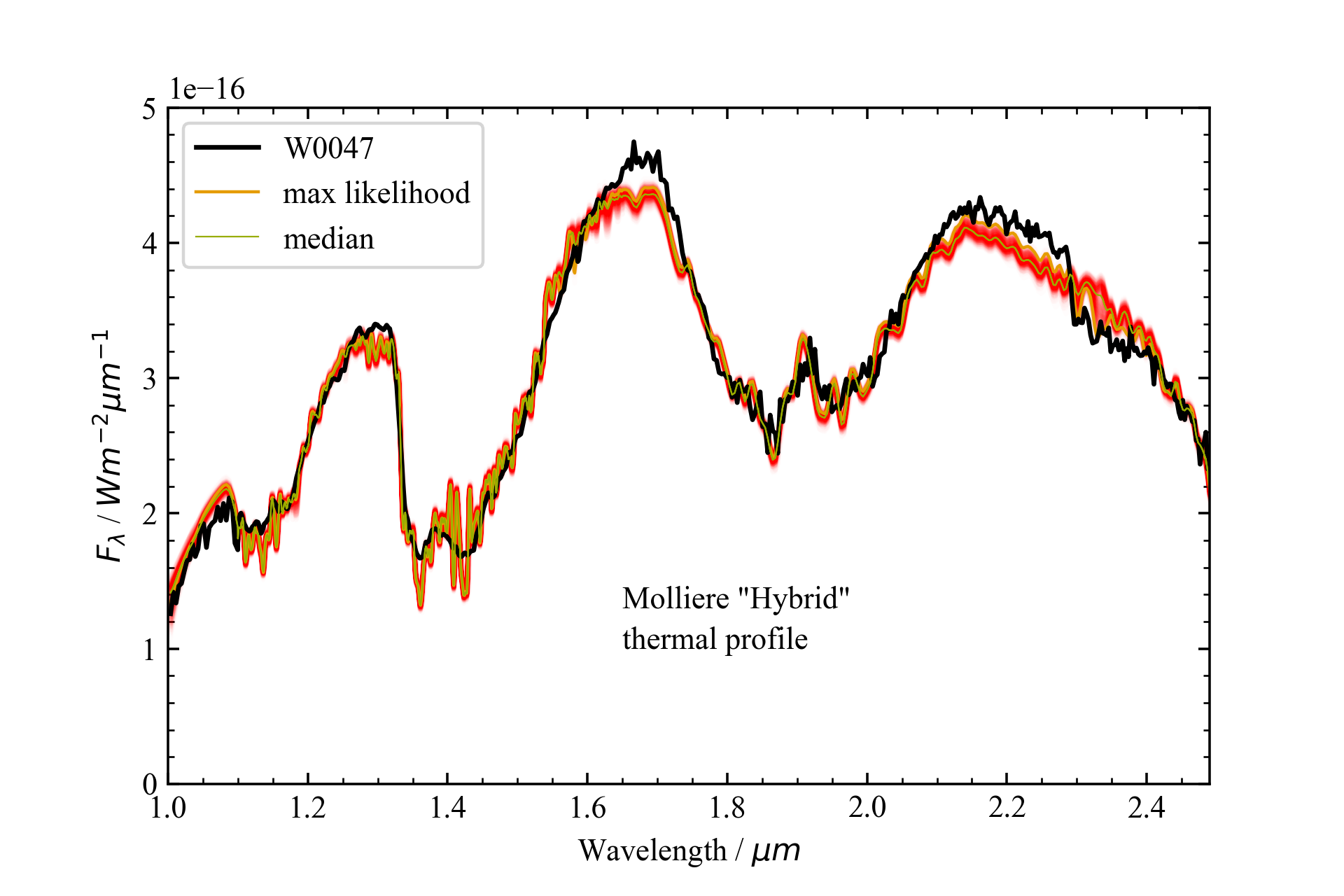}{0.50\textwidth}{\small(c)}}
    \caption{Retrieved forward model spectra for the power law deck models for \WISE. The maximum-likelihood spectrum is shown in orange and the median spectrum in green. The SpeX prism data is shown in black. (a) \WISE~retrieved spectra for the five point spline thermal profile, (b) Lavie profile, (c) Molli\`ere ``Hybrid" profile}
    \label{fig:retrieved_specta_W0047}
\end{figure}

For the retrieved spectra for \WISE~ (Figure \ref{fig:retrieved_specta_W0047}a), the five point spline approach underestimates the flux in the $H$ and $K$ band peaks hence the CO bands ($\approx$ 2.2 -- 2.4 $\mu$m) are not fit well. This is a similar conclusion to what was found for \BD. Figure \ref{fig:retrieved_specta_W0047}b shows the retrieved spectra for \WISE~for the Lavie thermal profile.  The retrieved spectrum generally traces the observed spectra and has a better fit to the CO feature in the $K$ band and does not fit the flux peaks of the triangular $H$ band region as with \BD. Similar to the five point spline profile the Molli\`ere  thermal profile retrieved spectra underestimates and poorly fits the flux in the $H$ and $K$ bands. For all three thermal parameterizations, the retrieved spectra underestimates the flux peak in the $J$-band. 
\par
\begin{figure}[H]
\centering
\gridline{\fig{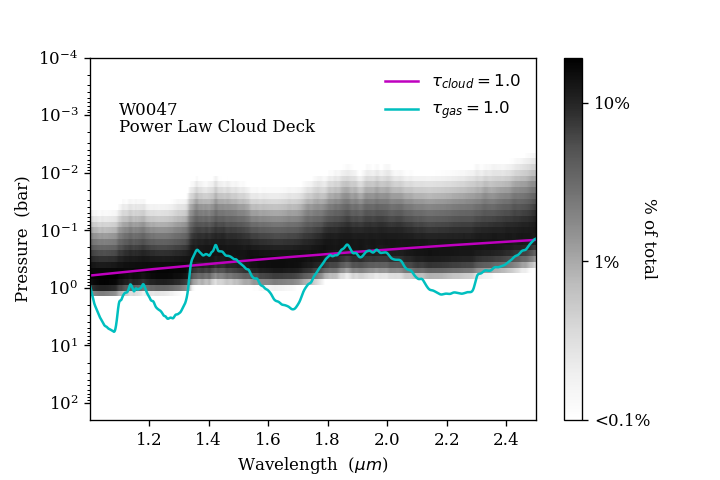}{0.50\textwidth}{\small(a)}}
     \gridline{\fig{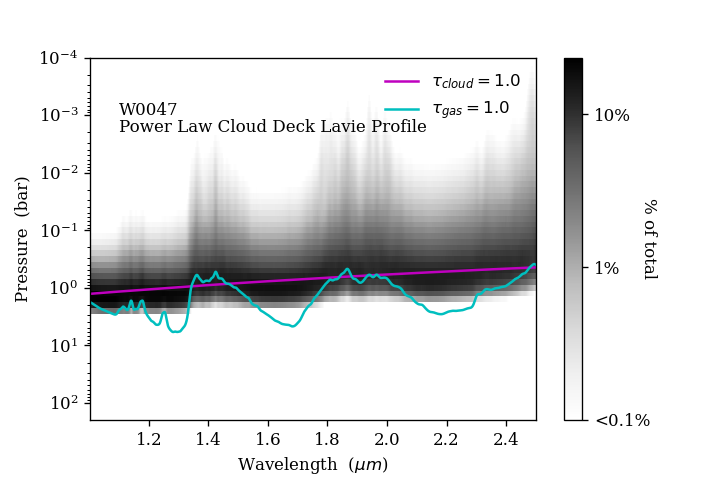}{0.50\textwidth}{\small(b)}}
     \gridline{\fig{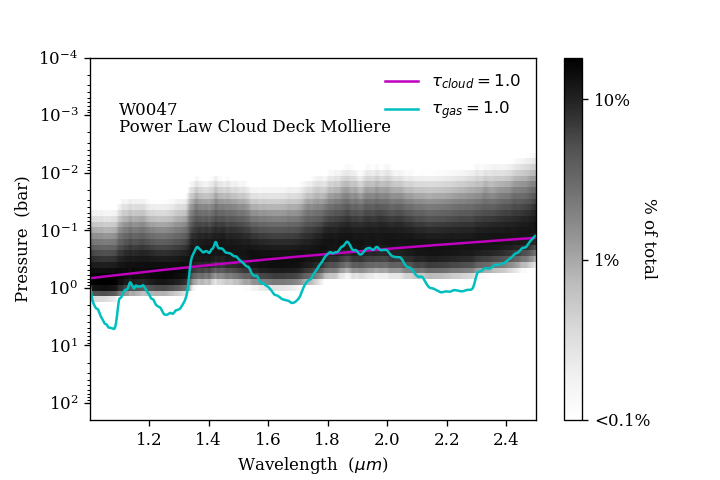}{0.50\textwidth}{\small(c)}}
 \caption{Contribution functions for \WISE~based on top-ranked model for each thermal parameterization cases. The black shading shows the percentage contribution of each pressure to the flux at each wavelength. $\tau=1$ lines are included for gas phase opacities (cyan),  and a cloud (magenta) (a) \WISE~ contribution function for five point spline thermal profile, (b) Lavie profile, (c) Molli\`ere ``Hybrid" profile}
\label{fig:ontribution_functions_W0047}
\vspace{0.5cm}
\end{figure}

Figure \ref{fig:ontribution_functions_W0047} shows the contribution functions for \WISE~for the winning cloud deck power law models. For the five point spline thermal profile, the bulk of the flux between $\approx$ 10$^{-2}$ and 1 bar contributes to the observed spectrum. Across the spectrum for the five point spline thermal profile (Figure \ref{fig:ontribution_functions_W0047}a), the cloud opacity dominates, except $\approx$ 1.4 $\mu$m region, where H$_{2}$O opacity contributes significantly.  For the Lavie thermal profile, the bulk of the flux between $\approx$ 0.003 and 10 bars contributes to the observed spectrum. The cloud opacity has a large contribution across the full spectrum with the exception of a small region in the H band ($\approx$ 1.8 -- 2.0 $\mu$m), where gas opacity becomes more important. Notably, the observable photosphere extends to the lowest pressures of the three thermal parameterizations that were tested.
\par
For the  Molli\`ere  thermal profile, the cloud opacity dominates the spectrum except for the $\approx$ 1.4 $\mu$m region where the gas opacity is dominant. The contribution function for the Molli\`ere  thermal profile is similar to that of the five point spline where the bulk of the flux between $\approx$ 10$^{-2}$ and 1 bar contributes to the observed spectrum.

\subsection{Retrieved Gas Abundances and Fundamental Parameters}

\begin{figure}[H]
    \centering
    \gridline{\fig{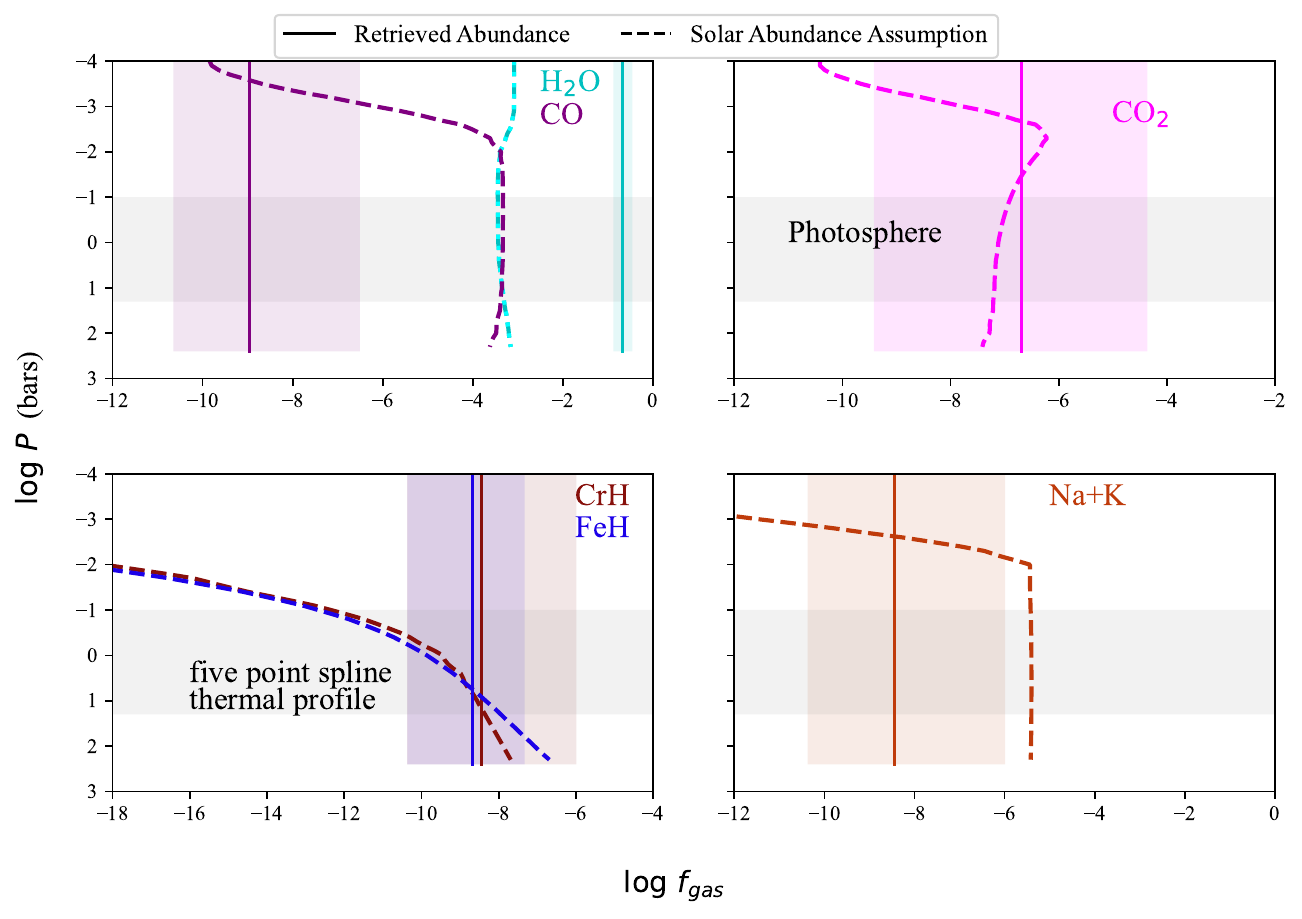}{0.49\textwidth}{\small(a)}}
     \gridline{\fig{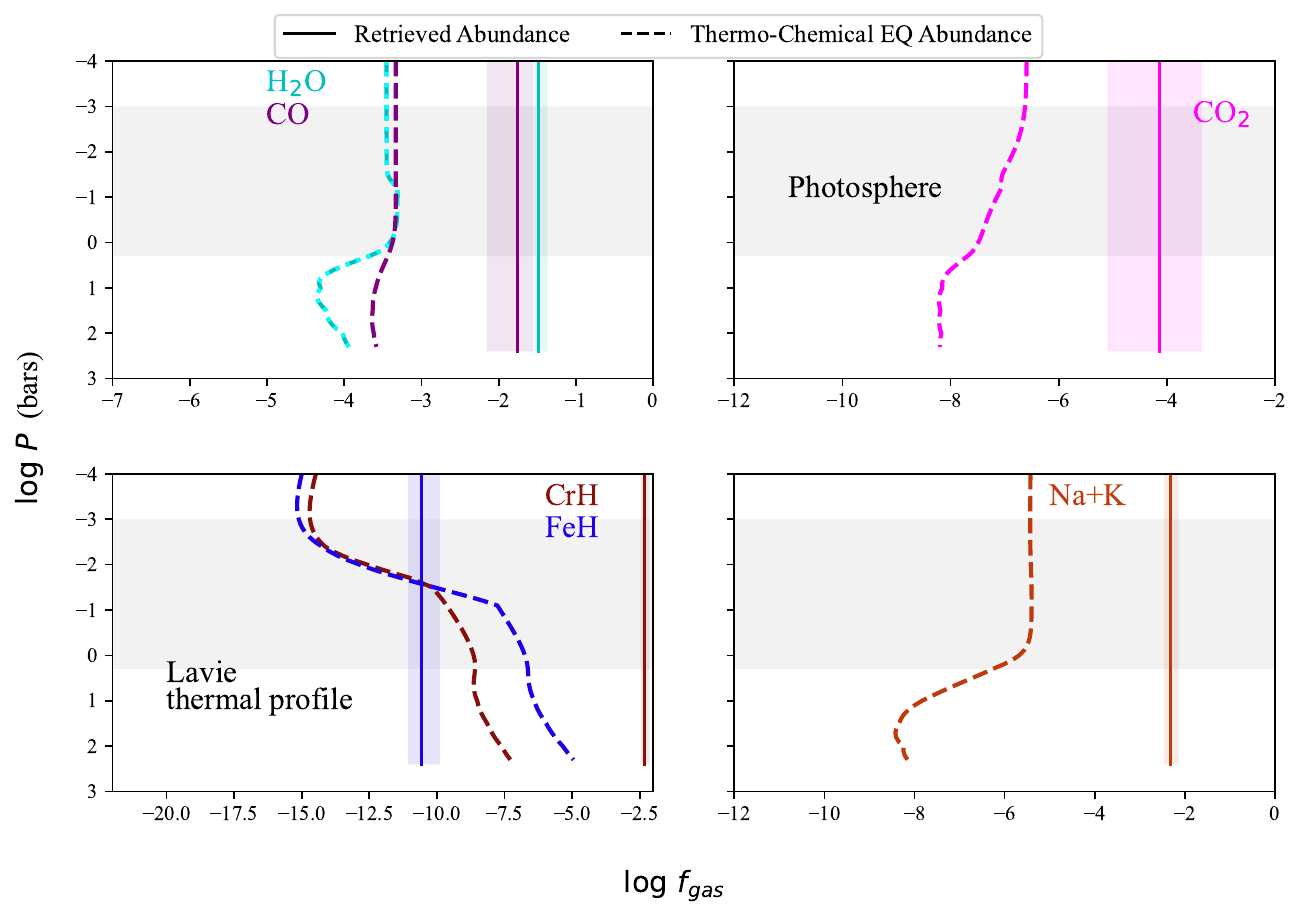}{0.49\textwidth}{\small(b)}}
     \gridline{\fig{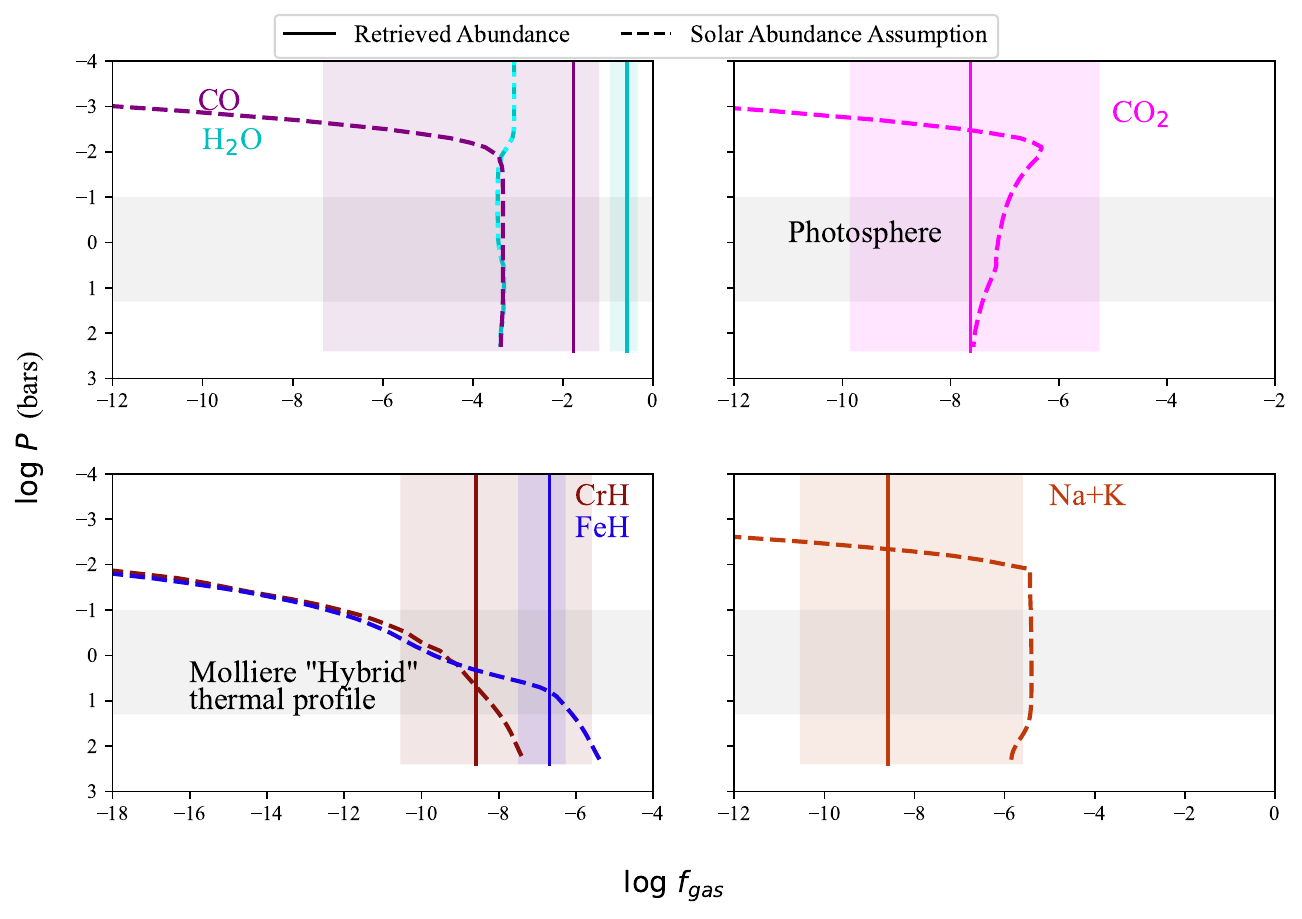}{0.49\textwidth}{\small(c)}}
    \caption{Retrieved uniform-with-altitude mixing abundances as compared to thermochemical equilibrium grid abundances for \WISE. Equilibrium predictions are shown as dashed lines. The solid straight lines and shading show our median retrieved values and 16th to 84th percentiles, respectively. (a)  five point spline thermal profile, (b) Lavie profile , (c) Molli\`ere ``Hybrid" profile}
    \label{fig:gas_abundances_W0047}
\end{figure}

Figure \ref{fig:gas_abundances_W0047} shows the retrieved gas mixing ratios compared to predictions from thermochemical equilibrium models for \WISE. As before, the thermochemical equilibrium models are calculated and interpolated for our derived metallicity and C/O ratio for the three thermal profile parameterizations. In the cases where the C/O ratio cannot be determined (due to implausibly high abundances) we assume solar values~\citep{Asplund2009}.

For the five point spline thermal model~(Figure \ref{fig:gas_abundances_W0047}a), the thermochemical equilibrium models predict many of the included gases to remain relatively constant with altitude (e.g., H$_{2}$O, CO, CO$_{2}$, Na+K) in the photosphere. However, FeH and CrH  are predicted to vary within the photosphere (denoted by the gray box). As with the retrieved abundances for \BD~for the five point spline, we retrieve an implausibly high H$_{2}$O abundance that does not match thermochemical equilibrium  predictions. Similarly, our retrieved CO abundance is depleted compared to predicted values. In Figure \ref{fig:corner_W0047_comparison}, we show the posterior distribution for retrieved and derived parameters for \WISE.

\par
We show the retrieved gas mixing ratios compared to the thermochemical equilibrium abundances for the Lavie profile~(Figure \ref{fig:gas_abundances_W0047}b). For this thermal profile parameterization, the retrieved abundances are more constrained, compared to the five point spline and Molli\`ere thermal profiles, however none of the retrieved abundances match the predicted thermochemical equilibrium abundances. 
\par
Similar to the five point spline thermal profile, for the Molli\`ere ``Hybrid" profile retrieved abundance (Figure \ref{fig:gas_abundances_W0047}c) we find an implausibly high H$_{2}$O abundance with respect to the predicted abundances assuming solar composition. While not a tight constraint, our CO abundance 1$\sigma$ values are within the limits of the model values.  FeH and CrH are predicted to vary with altitude, and show agreement with the retrieved values near the 10 bar region. Our retrieved Na+K values are not within the predicted range within the photosphere.

\subsection{Cloud Properties}

We find that the spectrum for \WISE~is best described by a cloudy model regardless of thermal profile parameterization. A summary of cloud parameters for \WISE~is shown in Table \ref{tab:cloud_properties}. The $\alpha$ parameter is positive for the power law, with the lowest $\alpha$ present for the Lavie profile. 
\par

The pressure of the cloud in the atmosphere of \WISE~is shown in various shaded regions to the right of Figure \ref{fig:thermal_profs_W0047}. The notable difference is that the pressure at which $\tau$ $>$ 1 for the Lavie profile is higher than those for the five point spline and  Molli\`ere ``Hybrid" thermal profiles, indicating a higher altitude cloud and shallower pressure of the corresponding contribution function (Figure \ref{fig:ontribution_functions_W0047}b).

\section{Discussion}
\label{sec:discussion}

 \subsection{The Complexity of Low Surface Gravity L-dwarfs}
Young L-dwarfs are redder at near-infrared wavelengths than field objects~\citep{Faherty2013,Faherty2016}, in part due to physical conditions brought on by a lower surface gravity. High altitude silicate clouds as a result of inefficient dust settling is a likely cause for low surface gravity object reddening~\citep{Looper2008,Suarez2022,Suarez2023}. The atmospheric physics and chemistry of low surface gravity L-dwarfs brings about changes to the features and near-infrared spectral shapes that can complicate modeling and analysis~\citep{Bonnefoy2013, Nowak2020,Whiteford2023}.

 \par
Complexities in the retrievals of young L-dwarfs like those seen within this work probably arise from compounding factors involving the emergence of condensate clouds. As noted by \cite{Rowland2023}, there exist three main challenges in the near-infrared modeling of L-dwarfs: (1) the limited range of the atmosphere that the near-infrared region can probe, which can limit our ability to constrain thermal profiles (2) the different opacity sources and absorbers which prefer uniform or non-uniform with altitude abundances, (3) the presence of clouds that can cause degeneracies with thermal profiles. 
\par

Notably, \cite{Burningham2021} found that chemical disequilibrium and non-solar elemental abundances are needed to accurately capture the complexity of cloudy L-dwarfs in future cloud models for retrievals. Related to this regime as well for the directly-imaged planets in the HR 8799 system, \cite{Lavie2017} commented that their future retrieval studies could also provide opportunities to explore disequilibrium chemistry through vertical mixing as well. 
\par

One clear conclusion from this work is that regardless of P-T profile chosen, both \BD~and \WISE~retrieval approaches strongly preferred the cloudy model.  However what was not clear was the parameterization of said cloud with the data on hand. 
It remains unclear how the interplay of clouds, equilibrium chemistry, and
gravity sensitive features  impact the spectra for red young L-dwarfs. Similarly it is unclear whether the full physical picture of all those parameters is fully represented within the \textit{Brewster} retrieval framework.  Hence performing the retrieval and achieving a physical or non-physical thermal profile and abundance output is itself a test on whether we have included the full range of needed physics in order to model these complex objects.

\subsection{``Free Retrievals'' vs Chemical Equilibrium Assumption}
Due to consistent, implausibly high abundances retrieved for \BD~and \WISE~we test our retrievals assuming thermochemical equilibrium, which has previously been demonstrated by \cite{Burningham2021} and \cite{Gonzales2022}. Instead of retrieving individual abundances, this method retrieves [M/H] and C/O. For simplicity, we test the chemical equilibrium assumption on the ``winning'' model for the Molli\`ere ``Hybrid'' profile for \BD. 
\par
In this model, we find the following parameters: [M/H] = 1.83$^{+0.10}_{-0.16}$,   C/O = 0.38$^{+0.16}_{-0.09}$, $R_\mathrm{Jup}$ = 0.51 $\pm$ 0.01 and $M_\mathrm{Jup}$ = 49.66$^{+16.78}_{-16.79}$. The chemical equilibrium assumption yields a smaller radius and higher mass than with our free retrieval. While the C/O and [M/H] might be more realistic under these assumptions our models prefer the free retrieval assumption over the chemical equilibrium assumption; $\Delta$logEv = 0 and 6.74 respectively.  These results are consistent with \cite{Burningham2021}, who also found that free retrievals were preferred for a higher gravity, cloudy L dwarf.

For the chemical equilibrium test (Figure \ref{fig:chemical_eq_plot_BD60B}a), we find that the retrieved P-T profile is not isothermal in the photosphere unlike the ``free'' retrieval runs. In the photosphere, the retrieved  P-T profile closely traces the grid model with fsed = 2. The 1$\sigma$ median cloud is within the observable photosphere. Notably, we find that the chemical equilibrium model provides a better fit to the $K$-band region containing the CO band head. 

\par
We find that while the chemical equilibrium model does not provide a statistically better fit ($\Delta$logEv) for \BD, it does provide a more plausible thermal profile. As we find that the chemical-equilibrium model (which allows abundances to vary, but in a fixed way) improves certain aspects of the retrieval; incorporating vertically-varying abundances into free-retrievals with \textit{Brewster} can prove to be a promising next step to take~\citep{Rowland2023}.

\subsection{Comparison with SED-derived Fundamental Parameters}
\label{sec:discuss_sed_comparision}
We find discrepancies for the fundamental parameters derived and extrapolated for both \BD~and~\WISE~regardless of thermal model parameterization compared to our SED-derived values. Our retrieved radii for \BD~and~\WISE~suffer from the ``small-radius problem" known to be present in retrieval analysis for L and L/T transition objects~\citep{Kitzman2020, Molliere2020, Gonzales2020, Gonzales2021,Gonzales2022, Burningham2021, Luber2022, Vos2023}. The retrieved radii for \WISE~were consistently $\sim$ 0.5$R_\mathrm{{Jup}}$ regardless of the thermal profile parameterization used. For \BD, the Molli\`ere ``Hybrid" profile has an upper limit of 0.86 $R_\mathrm{{Jup}}$, the highest of the derived radii, but still inconsistent with the SED  semi-empirical evolutionary model of 1.29 $\pm$ 0.06$R_\mathrm{{Jup}}$.  The high surface gravity for both \BD~and \WISE~is likely driven by the unrealistically small radii. The ``small radii" problem is predominantly seen in L dwarfs therefore it may be linked to poorly parameterized cloud properties given that condensates are a dominating and difficult to model component of their spectra. The high surface gravity is persistent regardless of the thermal profile tested -- also likely due to degeneracies between radius and surface gravity or a known issue with metallicity as discussed in \cite{Gonzales2021}
\par
Across the board of thermal parameterizations, the retrieved mass for \BD~is larger than the SED semi-empirical analysis (13.47 $\pm$ 5.3 $M_{\rm Jup}$). The Molli\`ere ``Hybrid" profile produced the lowest mass of (36.54$^{+13.94}_{-12.61}$  $M_{\rm Jup}$), however, this is more than 5$\sigma$ larger than the SED derived mass. A similar observance is found for~\WISE. This is likely due to the ``small'' radius problem driving a large surface gravity and in turn forcing a larger mass.
\par
Our derived $T_\mathrm{eff}$ is consistently lower than the SED derived values for both \BD~and \WISE. Typically the ``small radius problem" tends to push $T_\mathrm{eff}$ higher - whereas we see lower $T_\mathrm{eff}$ in both \BD~and~\WISE. Our low-luminosity could be due to our high retrieved H$_{2}$O abundance which in turn could be blocking flux for both \BD~and \WISE.

\begin{deluxetable}{lllll}[H]
\tabletypesize{\small}
\tablecolumns{10}
\tablewidth{0pt}
\setlength{\tabcolsep}{0.02in}
\tablecaption{Comparison of Fundamental and Derived Parameters for \BD~and \WISE~winning models for each thermal profile parameterizaton}
\label{tab:fund_derived_para_compar}
\tablehead{
\colhead{\textbf{\BD}}}
\startdata
\colhead{}&\colhead{\textbf{five point}}&\colhead{\textbf{Lavie}}&\colhead{\textbf{ Molli\`ere}}
 &\colhead{\textbf{SED}$^{a}$}\\
\hline
Radius ($R_{\rm Jup}$)& 0.59$^{+0.10}_{-0.06}$&0.51$\pm$0.01&0.75$^{+0.11}_{-0.08}$&$1.29\pm0.06$  \\
Mass ($M_{\rm Jup}$)&40.83$^{+10.47}_{-9.50}$&61.11$^{+9.53}_{-10.96}$&36.54$^{+13.94}_{-12.61}$&13.47 $\pm$  5.67\\
log $g$ (cm s$^{-2}$)&5.45$^{+0.11}_{-0.15}$&5.76$^{+0.061}_{-0.09}$&5.20$^{+0.16}_{-0.19}$&$4.26\pm0.20$\\
$T_{\rm eff}$ (K)&826.35$^{+42.10}_{-63.09}$&889.47$^{+7.19}_{-8.64}$&735.20$^{+45.62}_{-50.67}$&$1240\pm81$ \\
log ($L_{\rm bol}$ /L$_{\rm Sun}$)&--5.81$\pm$0.01&--5.81$\pm$0.00&--5.81$\pm$0.01&$-4.42\pm0.10$\\
C/O&---&0.02$^{+0.24}_{-0.02}$&---&---\\
$[M/H]$&2.71$^{+0.45}_{-0.38}$&--0.38$^{+0.35}_{-0.10}$&2.67$^{+0.48}_{-0.40}$&---\\
\hline
\colhead{\textbf{\WISE}}\\
\hline
Radius ($R_{\rm Jup}$)& 0.51$\pm$0.01&0.51$\pm$0.01 &0.51$\pm$0.01&$1.29\pm0.03$\\
Mass ($M_{\rm Jup}$)&50.28$^{+11.22}_{-10.43}$&59.54$^{+6.24}_{-3.29}$&45.45$^{+15.63}_{-23.39}$&$15.65\pm4.65$ \\
log $g$(cm s$^{-2}$)&5.68$^{+0.09}_{-0.10}$&5.75$^{+0.04}_{-0.03}$&5.63$^{+0.13}_{-0.30}$&$4.3\pm0.1$\\
$T_{\rm eff}$ (K)&872.35$^{+6.90}_{-9.01}$&875.86$^{+6.48}_{-7.56}$&871.47$^{+7.31}_{-10.44}$& $1291\pm51$\\
log ($L_{\rm bol}$ /L$_{\rm Sun}$)&--5.84$\pm$0.00&--5.84$\pm$0.00&--5.84$\pm$0.00&$-4.35 \pm0.06$\\
C/O&---&0.35$^{+0.12}_{-0.14}$&0.05$^{+0.28}_{-0.05}$&---\\
$[M/H]$&2.32$^{+0.30}_{-0.24}$&1.80$^{+0.13}_{-0.14}$&2.55$^{+0.31}_{-0.30}$&---\\
\enddata
\tablenotetext{a}{Values are from the fundamental parameters derived from SED analysis (see \S \ref{sec:SED}) }

\end{deluxetable}

\subsection{Host Star Chemistry Implications on the Secondary}
Recently, \cite{Calmari2024} introduced a theoretical chemical framework to predict cloud conditions of brown dwarfs based on FGK host stars chemical abundances. Using the approach of \cite{Calmari2024}, we utilized the chemical abundances of the host star BD+60 1417 discussed in Section \ref{sec:host_star_analysis} in order to assess the likely cloud composition of \BD.  Compared to other stars in the local solar neighborhood (e.g. \citealt{Brewer2016}) we find that BD+60 1417 has a relatively high Mg/Si ratio and a subsolar C/O ratio. The implications of this chemistry would be on the distribution of refractory elements into particular cloud phases (i.e. enstatite, forsterite, or quartz) and the percentage of oxygen into such silicate clouds.  According to \cite{Calmari2024}, using the BD+60 1417 elemental abundances of Mg, Si, C, and O we anticipate the formation of forsterite (Mg$_2$SiO$_4$) and enstatite (MgSiO$_3$) clouds in the atmosphere of the companion. If this companion is indeed relatively silicon depleted, we would not expect a best fit model containing a quartz cloud. As is shown in \cite{Burningham2021} and \cite{Vos2023}, we also expect to see a large portion of the Fe inventory to condense as a patchy Fe deck cloud under the silicate cloud layer. Additionally, based on our host star chemistry, we can estimate the percentage of oxygen sunk into these silicate clouds as done in \cite{Calmari2024}. For this metal-rich system, we calculate an oxygen sink percentage of approximately 24$\%$, which lies on the higher end of the local neighborhood distribution shown  in Figure 2 in \cite{Calmari2024}. In future modeling, we would expect this to have implications on retrieved versus bulk C/O ratio.

Finally, based on the results of \cite{Suarez2022}, we may also expect to see more enstatite than forsterite given that this source is also a low gravity late-type L dwarf. Consequently, a longer wavelength spectrum such as that enabled by JWST's MIRI spectrograph would yield strong insights into the atmosphere and ties to the host stars composition.

\subsection{Predictions for the Mid-Infrared Region}
\begin{figure}[H]
    \centering
    \includegraphics[width=1.1\columnwidth]{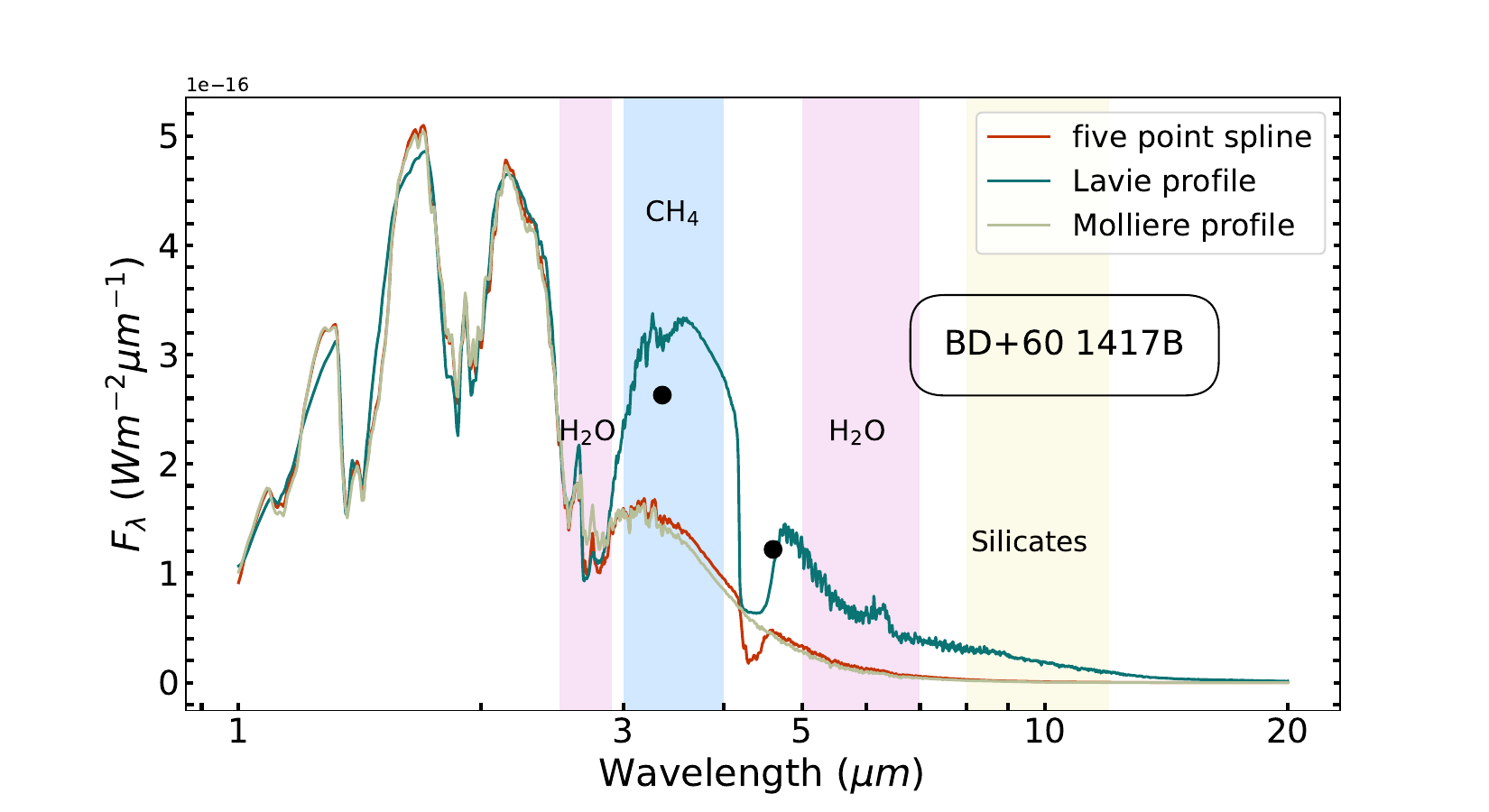}
\includegraphics[width=1.1\columnwidth]{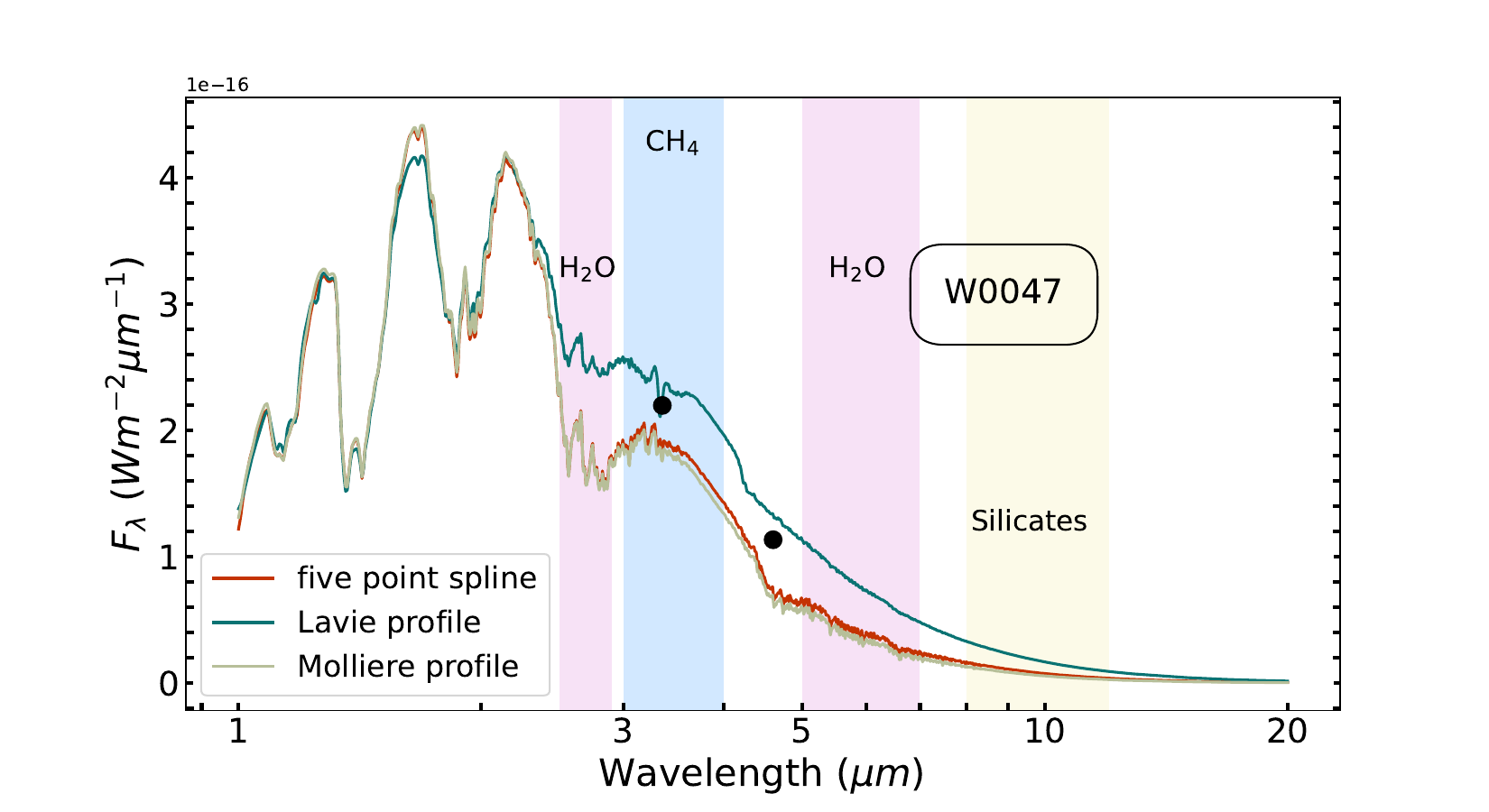}
\caption{Extracted mid-infrared spectra for \BD~and \WISE~with important molecular gas highlighted. } 
    \label{fig:extracted_spectra_BD60_W0047}
\end{figure}

With a limited number of planetary-mass worlds observed with JWST~\citep{Miles2022}, we have learned a wealth of information from their extended mid-infrared coverage, enabled by the MIRI coverage. In an effort to understand the mid-infrared region we utilize the best fit retrieved spectra from our three thermal profiles for both \BD~and \WISE~(Figure \ref{fig:retrieved_spectra_BD60B} \& Figure \ref{fig:retrieved_specta_W0047} respectively) to produce extended spectra predictions from 1 to 20 $\mu$m. As stated, while we can not constrain abundances with confidence in this work, we strongly predict the presence of clouds in both objects. We take the parameters for the max-likelihood in each best fit retrieval for a given P-T profile and rerun our forward model using the parameters over a wider wavelength range. The extrapolated spectra for \BD~and \WISE~are shown in Figure \ref{fig:extracted_spectra_BD60_W0047} and serve as predictive power for what might be seen in a JWST MIRI spectrum.
\par
We examine the extended retrieved spectrum relative to available photometric points for each source to compare/contrast the best P-T profile predictions. In the case of \BD~we find that for both the five point spline and the Molli\`ere ``Hybrid'' profiles, the predicted extended spectra underestimate the flux from the WISE photometry measurements. The extrapolated spectrum for the  Lavie profile most closely traces the WISE photometry, particularly at the W2 photometric point. For \WISE, we find that the extracted spectra again underestimates the flux from WISE photometry measurements for both the five point spline and Molli\`ere ``Hybrid'' profiles. In contrast, the Lavie profile extracted spectrum appears to slightly overestimate the flux compared to the WISE photometry, with the expected of W1 photometric point.  While no single thermal profile prediction appears to match current data, it is clear that the JWST MIRI spectrum has the potential to differentiate the P-T profile and parameterize cloud properties.

\section{Summary and Conclusion}
\label{sec:summary_conclusion}
In this work we present an atmospheric retrieval analysis of two young red L-dwarfs with similar spectral morphology and fundamental parameters, \BD~and \WISE, using the \textit{Brewster} retrieval framework. We also present abundances for the K0 host star, BD+60 1417, from high resolution spectroscopy from PEPSI/LBT (Section \ref{sec:host_star_analysis}).
\par
With the PEPSI data, we constrain an iron abundance of 0.27 $\pm$ 0.03 dex. We constrain measurements of the C/O = 0.23 $\pm$ 0.12 and Mg/Si = 1.41 $\pm$ 0.19. Given these parameters, we find that forsterite and enstatite clouds are expected to form and not quartz clouds. 
\par
In our retrieval analysis we test a number of different models and thermal parameterizations (five point spline, Lavie profile and a Molli\`ere ``Hybrid'' profile)  and find that both \BD~and \WISE~are best described by a cloudy atmosphere regardless of the chosen P-T profile. Regardless of thermal profile parameterization, the retrieved profile was generally not in agreement with forward models for both \BD~and \WISE. We find that for both \BD~and \WISE~the retrieved spectrum from \textit{Brewster} had difficulty fitting the $K$-band region, which contains the temperature sensitive CO feature.
\par

Lastly, although we test three different thermal profile parameterizations we find that we are unable to reliably constrain the C/O ratio of these planetary-mass objects due to retrieving implausibly high abundances (e.g H$_{2}$O). Generally we find that our abundances are not in agreement with thermochemical equilibrium models. While we experience challenges in modeling these young low-gravity objects, we find that improvements can be made in terms of modeling approach as we discover that non-uniform with altitude abundances provide a more realistic thermal profile for \BD. This work continues to provide technical lessons in retrieval analysis of young cloudy L-dwarfs.

\par

JWST has the capabilities to provide wider wavelength coverage and simultaneous higher resolution and signal-to-noise observations. Wide wavelength coverage will allow us to probe the pressure-temperature profile from (0.1 - 1 bar) through mid-infrared all the way to 10 bar pressure with near-infrared coverage. Near-infrared and mid-infrared data is crucial to retrieve a reliable P-T profile for early-to-mid L dwarfs \citep{Burningham2021,Vos2023,Rowland2023}. Future retrievals of \BD~and \WISE~with JWST data would allow us to accurately constrain the P-T profile and cloud parameters to break degeneracies between clouds and composition. 
\facilities{Exoplanet Archive, LBT (PEPSI)}
\software{\texttt{Brewster} \citep{Burningham2017}, \texttt{PyMultiNest \citep{Buchner2014}}, \texttt{corner.py} \citep{Foreman-Mackey_2013}} 
 
\section*{Acknowledgements}

We thank our anonymous reviewer for providing detailed recommendations that significantly improved the quality of this paper. 

This research has made use of the NASA Exoplanet Archive, which is operated by the California Institute of Technology, under contract with the National Aeronautics and Space Administration under the Exoplanet Exploration Program.
\par
NASA's Astrophysics Data System Bibliographic Services together with the VizieR catalogue access tool and SIMBAD database operated at CDS, Strasbourg, France, were invaluable resources for this work. 
This publication makes use of data products from the Two Micron All Sky Survey, which is a joint project of the University of Massachusetts and the Infrared Processing and Analysis Center/California Institute of Technology, funded by the National Aeronautics and Space Administration and the National Science Foundation.

This work benefited from involvement in ExoExplorers, which is sponsored by the Exoplanets Program Analysis Group (ExoPAG) and NASA’s Exoplanet Exploration Program Office (ExEP).
Caprice Phillips thanks the LSSTC Data Science Fellowship Program, which is funded by LSSTC, NSF Cybertraining Grant $\#$1829740, the Brinson Foundation, and the Moore Foundation; her participation in the program has benefited this work. E.J.G. is supported by an NSF Astronomy and Astrophysics Postdoctoral Fellowship under award AST-2202135.
E.G. acknowledges support from the Heising-Simons Foundation for this research. J. M. V. acknowledges support from a Royal Society - Science Foundation Ireland University Research Fellowship (URF$\backslash$1$\backslash$221932). BB acknowledges support from UK
Research and Innovation Science and Technology Facilities Council [ST/X001091/1]. 

This work benefited from the 2023 Exoplanet Summer Program in the Other Worlds Laboratory (OWL) at the University of California, Santa Cruz, a program funded by the Heising-Simons Foundation and NASA.
\par
This work has made use of data from the European Space Agency (ESA) mission
{\it Gaia} (\url{https://www.cosmos.esa.int/gaia}), processed by the {\it Gaia}
 This publication makes use of data products from the Wide-field Infrared Survey Explorer, which is a joint project of the University of California, Los Angeles, and the Jet Propulsion Laboratory/California Institute of Technology, funded by the National Aeronautics and Space Administration.
 Data Processing and Analysis Consortium (DPAC,
\url{https://www.cosmos.esa.int/web/gaia/dpac/consortium}). Funding for the DPAC
has been provided by national institutions, in particular the institutions
participating in the {\it Gaia} Multilateral Agreement

This work is based on observations made with the Large Binocular Telescope. The LBT is an international collaboration among institutions in the United States, Italy and Germany. LBT Corporation partners are: The University of Arizona on behalf of the Arizona Board of Regents; Istituto Nazionale di Astrofisica, Italy; LBT Beteiligungsgesellschaft, Germany, representing the Max- Planck Society, The Leibniz Institute for Astrophysics Potsdam, and Heidelberg University; The Ohio State Uni- versity, representing OSU, University of Notre Dame, University of Minnesota and University of Virginia.

The data used in this publication were collected through the MENDEL high performance computing (HPC) cluster at the American Museum of Natural History. This HPC cluster was developed with National Science Foundation (NSF) Campus Cyberinfrastructure support through Award 1925590. This
work was also supported by the National Science Foundation via awards 1909776, 1614527 and AST-1909837.

Computation reported in this work was also carried out on the Unity cluster of the College of Arts and Sciences at the Ohio State University. The computational resources and support provided are gratefully acknowledged.

\restartappendixnumbering 
\appendix
\section{Chemical Equilibrium Assumption Model} 
\label{sec:appendix_chemicalequilibrium_test}

\begin{figure}[H]
     \gridline{\fig{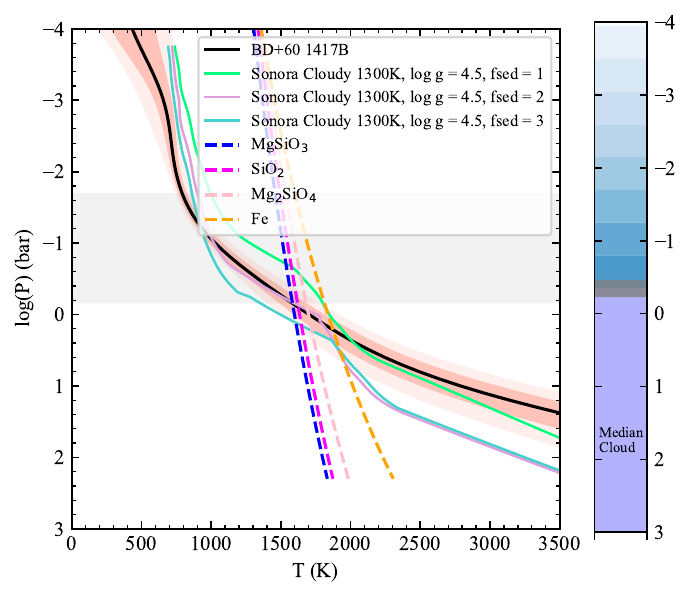}{0.5\textwidth}{\small(a)} \fig{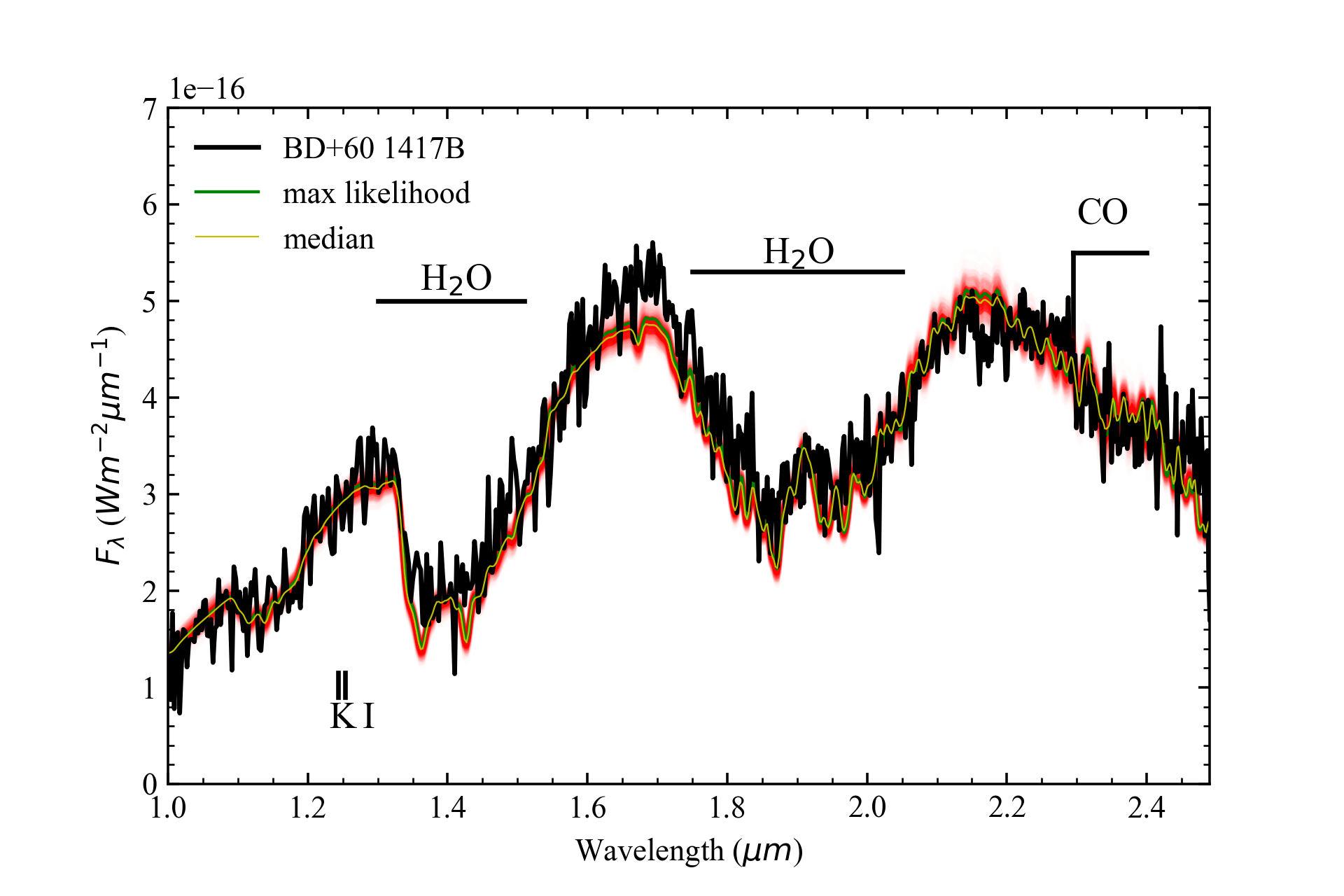}{0.6\textwidth}{\small(b)}} \gridline{\fig{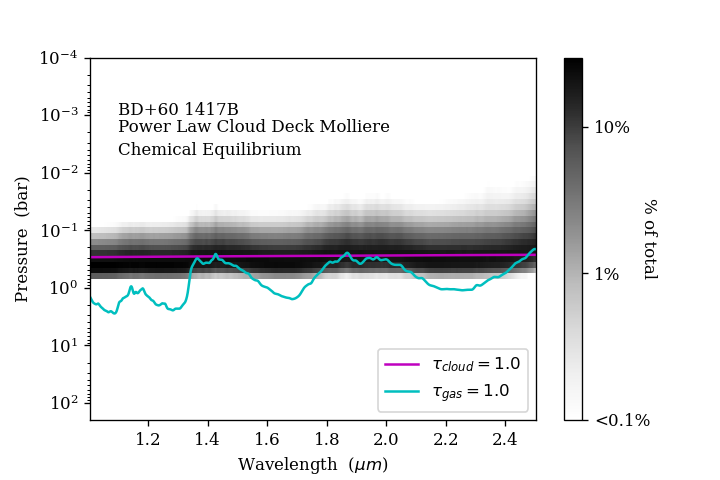}{0.6\textwidth}{\small(c)}}
   
    \caption{(a) retrieved thermal profile for the winning Molli\`ere ``Hybrid'' profile for \BD~under the assumption of chemical equilibrium. (b) retrieved forward model spectra for \BD~for chemical equilibrium (c) contribution function for \BD~for the winning model}
    \label{fig:chemical_eq_plot_BD60B}
\end{figure}

In Figure \ref{fig:chemical_eq_plot_BD60B}, we present the retrieval plots for the chemical equilibrium test for the winning Molli\`ere ``Hybrid'' profile for \BD. While this model is not preferred over the ``free'' retrieval we discuss the implications of testing this retrieval set up. 
\section{Corner Plots}
\label{sec:appendix_cornerplots}
\begin{figure}[H]
    \centering
    \includegraphics[width=\textwidth]{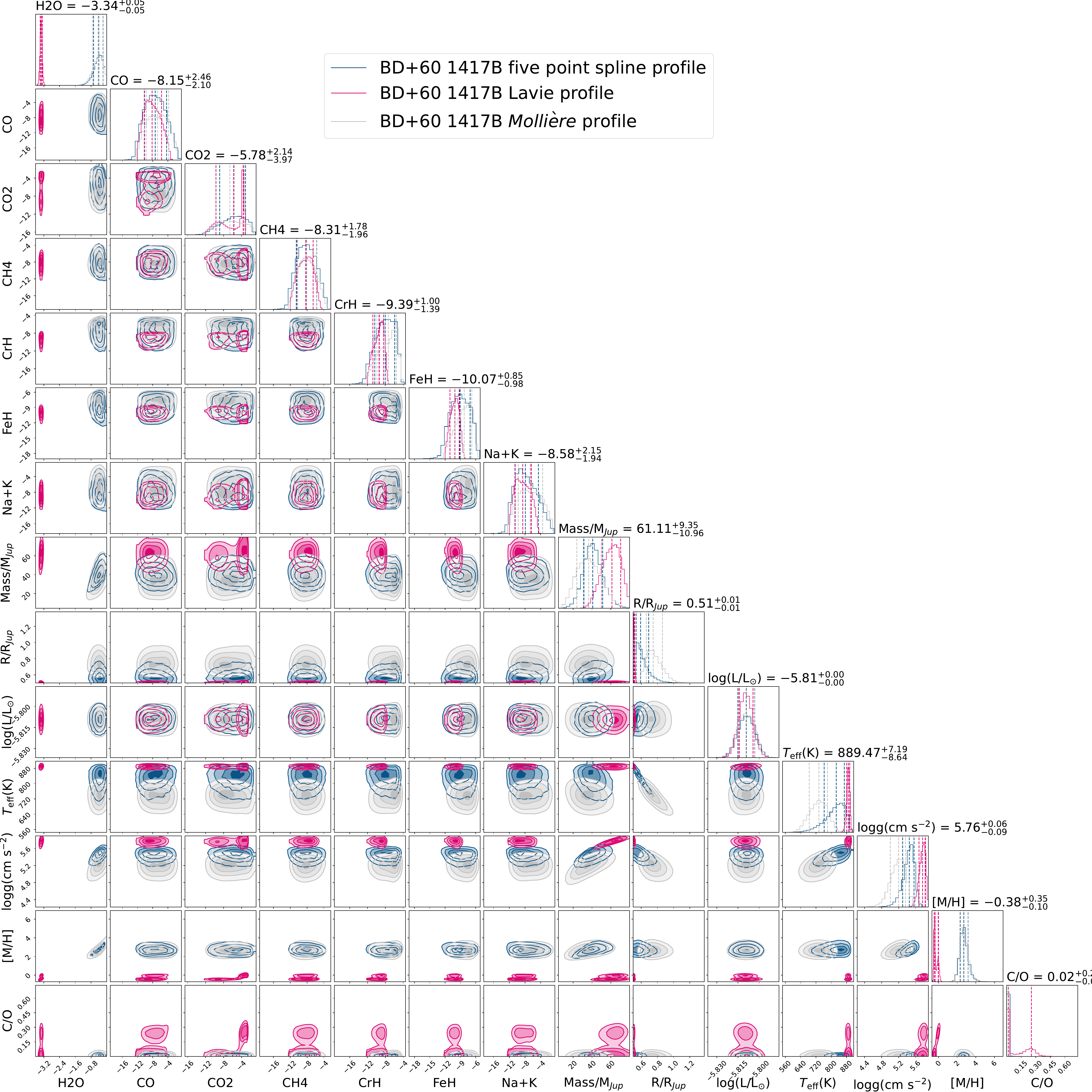}
    \caption{Best Fit Models for the five point spline thermal profile, Lavie profile, and Molli\`ere profile for \BD: power law cloud deck (blue), grey cloud deck (pink), and power law cloud deck (silver) for best fits respectively. Posterior probability distributions for retrieved gas abundances, mass, radius and derived quantities of the best fit models.  Far right diagonal plots show marginalized posteriors for each parameter along with 2D parameter correlation histograms. Dashed lines show the median retrieved value
(reported value) along with 1$\sigma$ confidence intervals. Gas abundances are in units of dex.}
\label{fig:corner_BD60_comparison}
\end{figure}

\begin{figure}[H]
    \centering
    \includegraphics[width=\textwidth]{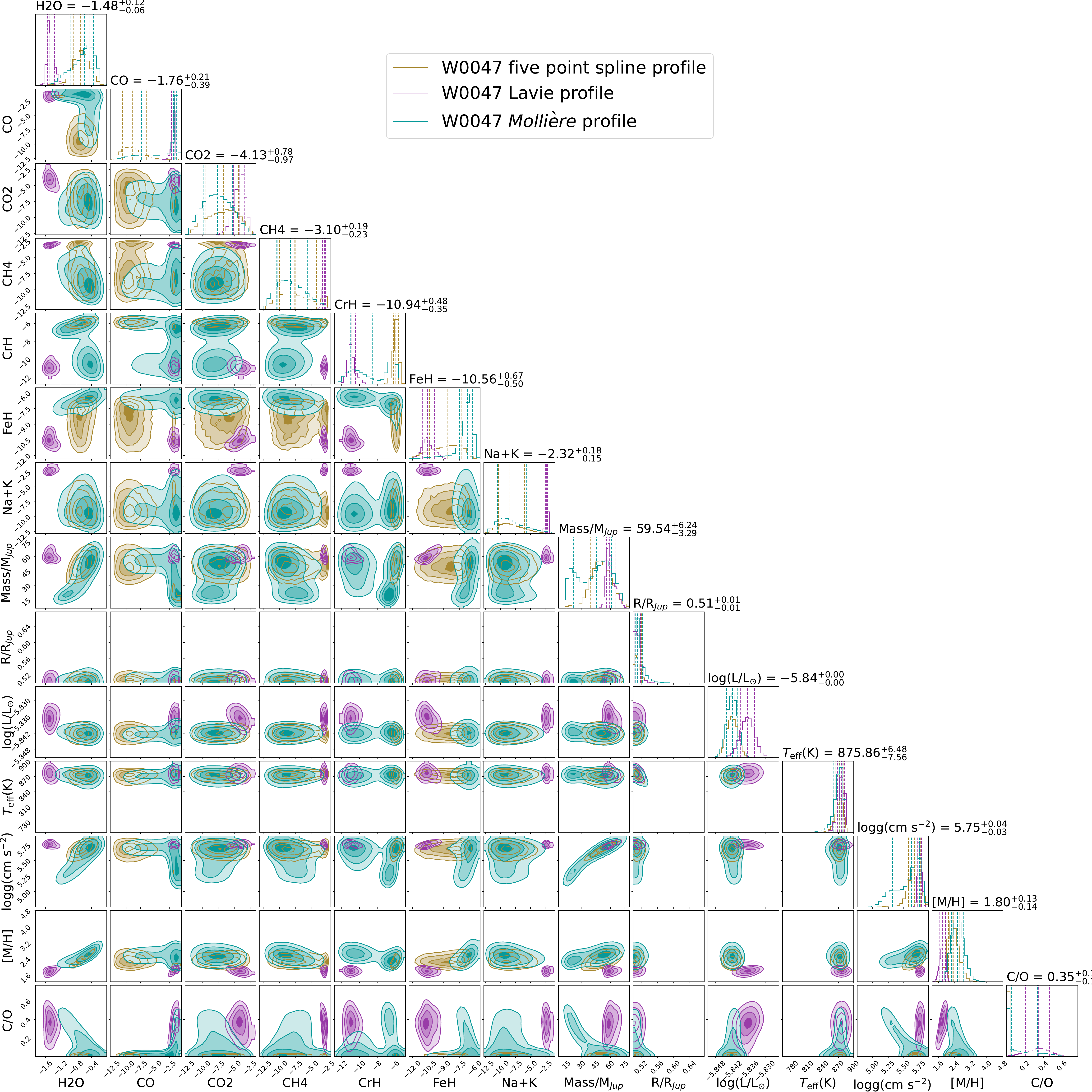}
    \caption{Best Fit Models for the five point spline thermal profile, Lavie profile, and Molli\`ere profile for \WISE: power law cloud deck (gold), power law cloud deck (purple), and power law cloud deck (green) for best fits respectively. Posterior probability distributions for retrieved gas abundances, mass, radius and derived quantities of the best fit models.  Far right diagonal plots show marginalized posteriors for each parameter along with 2D parameter correlation histograms. Dashed lines show the median retrieved value
(reported value) along with 1$\sigma$ confidence intervals. Gas abundances are in units of dex.}
\label{fig:corner_W0047_comparison}
\end{figure}

In Figure \ref{fig:corner_BD60_comparison} and \ref{fig:corner_W0047_comparison}, we present the posteriors for the gas abundances and derived fundamental parameters for the three different thermal profile parameterizations used in our retrievals for \BD~and \WISE~respectively.

\clearpage
\bibliography{sample631}{}
\bibliographystyle{aasjournal}

\end{CJK*}

\end{document}